\documentclass[aps,twocolumn]{revtex4-1}

\usepackage{times}

\usepackage{amsmath}
\usepackage{amsfonts}
\usepackage{amssymb}
\usepackage{epsfig}

\begin{document}

\title{Electron quantum optics in ballistic chiral conductors}

\author
{E. Bocquillon$^{1}$ $^\dag$, V. Freulon$^{1}$, F.D. Parmentier $^1$, J.-M Berroir$^{1}$,  B. Pla\c{c}ais$^{1}$, C. Wahl $^{2}$, J. Rech $^{2}$, T. Jonckheere $^{2}$, T. Martin $^{2}$, C. Grenier $^{3}$, D. Ferraro $^{3}$, P. Degiovanni$^{3}$, and G. F{\`e}ve $^{1}$$^\ast$ \\
\medskip
\normalsize{$^{1}$ Laboratoire Pierre Aigrain, Ecole Normale Sup\'erieure, CNRS (UMR 8551), Universit\'e Pierre et Marie Curie, Universit\'e Paris Diderot, 24 rue Lhomond, 75231 Paris Cedex 05}\\
\normalsize{$^{2}$Aix-Marseille Universit\'e, CNRS, CPT, UMR 7332, 13288 Marseille, France.
Universit\'e de Toulon, CNRS, CPT, UMR 7332, 83957 La Garde, France.}\\
\normalsize{$^{3}$Universit\'e de Lyon, F\'ed\'eration de Physique Andr\'e Marie Amp\`ere,
CNRS - Laboratoire de Physique de l'Ecole Normale Sup\'erieure de Lyon
46 All\'ee d'Italie, 69364 Lyon Cedex 07,France.} \\
\normalsize{$^\dag$ Currently at Physikalisches Institut (EP3), University of W\"{u}rzburg, Am Hubland, D-97074 W\"{u}rzburg, Germany.}\\
\normalsize{$^\ast$ To whom correspondence should be addressed;
E-mail:  feve@lpa.ens.fr.} }

%\author{E. Bocquillon $^{1}$, V. Freulon$^{1}$, F.D. Parmentier $^{1}$, J.-M Berroir$^{1}$, B. Plaçais$^{1}$, C. Wahl$^{2}$, J. Rech$^{2}$, T. Jonckheere$^{2}$, T. Martin$^{2}$, C. Grenier$^{3}$, D. Ferraro$^{3}$, P. Degiovanni$^{3}$, G. F\`{e}ve $^{1}$
%\\
%\normalsize{ $^1$ Laboratoire Pierre Aigrain, Ecole Normale Sup\'erieure, CNRS (UMR 8551), Universit\'e Pierre et Marie Curie, Universit\'e Paris Diderot} \\
%\normalsize{24 rue Lhomond, 75231 Paris Cedex 05}\\
%\normalsize{$^2$ Aix-Marseille Universit\'e, CNRS, CPT, UMR 7332, 13288 Marseille, France}\\
%\normalsize{$^3$ Universit\'e de Lyon, F\'ed\'eration de Physique Andr\'e Marie Amp\`ere,
%CNRS - Laboratoire de Physique de l'Ecole Normale Sup\'erieure de Lyon}\\
%\normalsize{ 46 All\'ee d'Italie, 69364 Lyon Cedex 07,France.} }

\begin{abstract}
The edge channels of the quantum Hall effect provide one dimensional chiral and ballistic wires along which electrons can be guided in optics like setup. Electronic propagation can then be analyzed using concepts and tools derived from optics. After a brief review of electron optics experiments performed using stationary current sources which continuously emit electrons in the conductor, this paper focuses on triggered sources, which can generate on-demand a single particle state. It first outlines the electron optics formalism and its analogies and differences with photon optics and then turns to the presentation of single electron emitters and their characterization through the measurements of the average electrical current and its correlations. This is followed by a discussion of electron quantum optics experiments in the Hanbury-Brown and Twiss geometry where two-particle interferences occur. Finally, Coulomb interactions effects and their influence on single electron states are considered.
\end{abstract}

\maketitle
% \noindent
\section*{Introduction}

Mesoscopic electronic transport aims at revealing and studying the quantum mechanical effects that take place in micronic samples, whose size becomes shorter than the coherence length on which the phase of the electronic wavefunction is preserved at very low temperatures. In particular, such effects can be emphasized when the electronic propagation in the sample is not only coherent but also ballistic and one-dimensional. The wave nature of electronic propagation then bears strong analogies with the propagation of photons in vacuum. Using analogs of beam-splitters and optical fibers, the electronic equivalents of optical setups can be implemented in a solid state system and used to characterize electronic sources. These optical experiments provide a powerful tool to improve the understanding of electron propagation in quantum conductors. Inspired by the controlled manipulations of the quantum state of light, the recent development of single electron emitters has opened the way to the controlled  preparation, manipulation and characterization of single to few electronic excitations that propagate in optics-like setups. These electron quantum optics experiments enable to bring quantum coherent electronics down to the single particle scale. However, these experiments go beyond the simple transposition of optics concepts in electronics as several major differences occur between electron and photons. Firstly statistics differ, electrons being fermions while photons are bosons. The other major differences come from the presence of the Fermi sea and the Coulomb interaction. While photon propagation is interaction free in vacuum, electrons propagate in the sea of the surrounding electrons interacting with each others through the long range Coulomb interaction turning electron quantum optics into a complex many body problem.

This article will be restricted to the implementation of such experiments in Gallium Arsenide two-dimensional electron gases. These samples provide the high mobility necessary to reach the ballistic regime and by applying a high magnetic field perpendicular to the sample enable to reach the quantum Hall effect in which electronic propagation occurs along one dimensional chiral edge channels. The latter situation is the most suitable to implement electron optics experiments. Firstly because electrons can be guided along one dimensional quantum rails, secondly because chirality prevents interferences between the electron sources and the optics-like setup used to characterize it. After briefly recalling the main analogies between electron propagation along the one dimensional chiral edge channels and photon propagation in optics setups, we will review the pioneer experiments that have been realized in these systems and that demonstrate the relevance of these analogies. Most of these experiments have been realized with DC sources that generate a continuous flow of electrons in the system and thus do not reach the single particle scale. The core of this review will then deal with the generation and characterization of single particle states using single electron emitters.

\subsection{Optics-like setups for electrons propagating along one dimensional chiral edge channels}

The first ingredient to implement quantum optics experiment with electrons is a medium in which ballistic and coherent propagation is ensured on a large scale. In condensed matter, this is provided by two-dimensional electron gases: these semiconductor hetero-structures (in our case and most frequently GaAs-AlGaAs) are grown by molecular beam epitaxy, which supplies crystalline structures with an extreme degree of purity. Thus mobilities up to about $10-30\times 10^6\ \rm{cm^2.V^{-1}.s^{-1}}$ have been reported \cite{Rossler2011,Hatke2012,Lin2012}, and mean-free path $l_e$ as well as phase coherence lengths $l_\phi$ can be on the order of $10-20\,\mu$m.
These properties enable to pattern samples with e-beam lithography in such a way that the phase coherence of the wavefunction is preserved over the whole structure, thus fulfilling a first requirement to build an electron optics experiment in a condensed matter system. The simplest interference pattern can be produced for example in Young's double-slit experiment \cite{Schuster1997} where the phase difference between paths is tuned via the enclosed Aharonov-Bohm flux, leading to the observation of an interference pattern in the current.

\begin{figure}[h!]
\centering\includegraphics[width=\columnwidth]{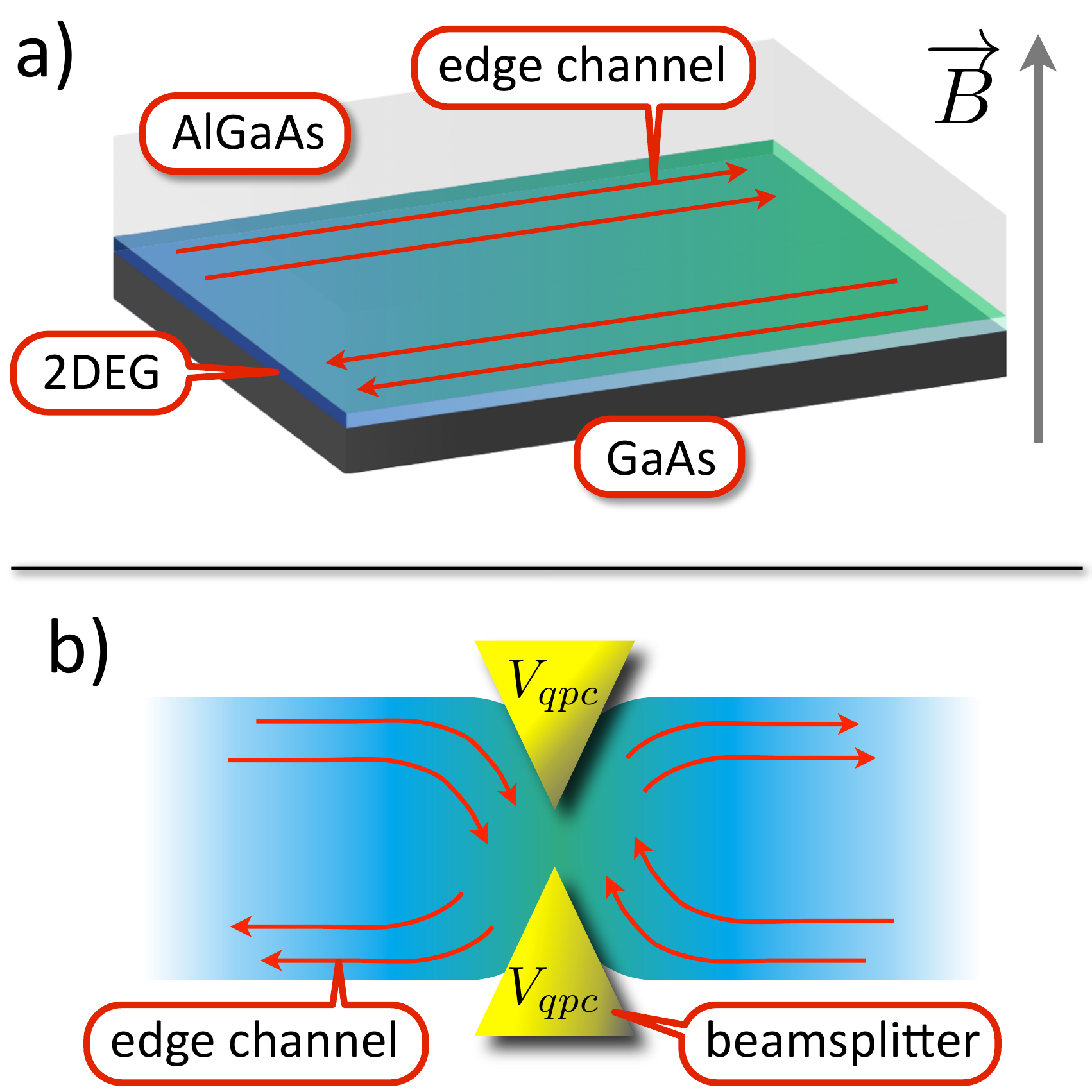}
\caption{a) Schematics of a 2DEG in the integer quantum Hall regime : when a strong perpendicular magnetic field is applied, electronic transport is governed by chiral edge channels. b) Schematics of a quantum point contact (QPC): when a negative voltage $V_{qpc}$ is applied on split gates deposited above the 2DEG, a tunnel barrier is created and enables to realize the electronic analog of a beamsplitter.}\label{fig:EdgeChannelQPC}
\end{figure}

Besides, electrons have to be guided from their emission to their detection through all the optical elements. A powerful implementation of phase coherent quantum rails is provided by (integer) quantum Hall effect. Under a strong perpendicular magnetic field, electronic transport in the 2DEG is governed by chiral one-dimensional conduction channels appearing on the edges while the bulk remains insulating (see Fig.\ref{fig:EdgeChannelQPC}a). The appearance of such edge channels results from the bending of Zeeman-split Landau levels near the edges of the samples \cite{Yoshioka2002}. Importantly, these edge channels are chiral: electrons flow with opposite velocities on opposite edges. The number of filled landau levels called the filling factor $\nu$ is the number of one dimensional channels flowing on one edge. It depends on the magnetic field: as $B$ increases, the Landau levels are shifted upward with respect to the Fermi energy, so that the number of Zeeman-split Landau levels crossing the Fermi level (that is, the number of filled Landau levels) decreases. The conductance $G$ of the 2DEG is quantized in units of the inverse of the Klitzing resistance $e^2/h=R_K^{-1}$ (where $R_K=25.8\, \rm{k\Omega}$) and given by the number of edge channels: $G=\nu R_K^{-1}$. Many experiments are performed at filling factor $\nu=2$, where electronic transport occurs on two edge channels, which are spin-polarized, corresponding to two Zeeman-split levels.
In the quantum Hall regime, the mean free path of electrons is considerably increased, up to $l_e\sim100\,\mu$m: the chirality imposed by the magnetic field reduces backscattering drastically, as an electron has to scatter from one edge to the counter-propagating one to backscatter, which can only be done when Landau levels are partially filled in the bulk. Beside the absence of backscattering in the edge channels \cite{Buttiker1988}, large phase coherence lengths have also been measured ($l_\phi\sim20\,\mu$m at 20 mK \cite{Roulleau2008}). However, backscattering can be induced locally on a controlled way using a quantum point contact (QPC) which consists of a pair of electrostatic gates deposited on the surface of the sample with a typical distance between the gates of a few hundreds of nanometers. The typical geometry of QPC gates is shown in Fig.\ref{fig:EdgeChannelQPC} b): when a negative gate voltage $V_{qpc}$ is applied on the gates, a constriction is created in the 2DEG between the gates because of electrostatic repulsion. This constriction gives rise to a potential barrier, the shape of which can be determined from the geometry of the gates \cite{Buttiker1990}. At high magnetic field, the transmission through the QPC is described in terms of edge channels following equipotential lines, which are reflected one by one as the QPC gate voltage is swept towards large negative values \cite{VanWees1988}. The conductance at low magnetic fields presents steps in units of $2e^2/h$ as Landau levels are spin-degenerate. At higher magnetic field, the height of the conductance steps is equal to $e^2/h$, reflecting Zeeman-split Landau levels and spin-polarized edge channels. Between two conductance plateaus, one of the edge channels is partially transmitted and  accounts for a contribution $T\frac{e^2}{h}$ to the conductance, proportional to the transmission probability $T$. In particular, when set at the exact half of the opening of the first conductance channel, the outer edge channel is partially transmitted with a probability $T=0.5$, while all other edge channels are fully reflected. The quantum point contact therefore acts as a tunable, channel-selective electronic beamsplitter in full analogy with the beamsplitters used in optical setups.

In the quantum Hall effect regime, electrons thus propagate along one-dimensional ballistic and phase coherent chiral edge channels which can be partitioned by electronic beamsplitters. These are the key ingredients to implement optics-like setups in electronics. The last missing elements are the electronic source that emits electrons and the detection apparatus. The measurement of light intensity and its correlations in usual quantum optics experiments is replaced by the measurement of the electrical current and its fluctuations (noise) for electrons. Concerning the electron emitter, this review will focus on triggered emitters that can emit particles on demand in the conductor. However, most of electron optics experiments and in particular the first ones have been performed using stationary dc sources that generate a continuous flow of charges in the system. Such a source can be implemented by applying a voltage bias $V$ to the edge channel, hence shifting the chemical potential of the edge by $-e V$. As a result, electrons generated in the edge channel are naturally regularly ordered, with an average time $h/eV$ between charges \cite{Martin1992}. The origin of this behavior is Pauli's exclusion principle, that prevents the presence of two electrons at the same position in the electron beam. As a consequence of Fermi statistics, a voltage biased ballistic conductor naturally produces a noiseless current \cite{Reznikov1995,Kumar1996}. Starting from the late nineties, many electron optics experiments have been performed to investigate the coherence and statistical properties of such sources.

\subsection{Electron optics experiments} \label{elecopticsexp}

The coherence properties of  stationary electron sources have been studied in  electronic Mach-Zehnder interferometers \cite{Ji2003,Litvin2007,Roulleau2008,Bieri2009}. Using two QPC's as electronic beamsplitters and benefiting from the ballistic propagation of electrons along the edges, single electron interferences can be observed in the current flowing at the output of the interferometer. The phase difference between both arms can be varied by electrostatic influence of an additional gate or by changing the magnetic field, thus changing the magnetic flux in the closed loop of the interferometer. This constitutes a very striking demonstration of the phase coherence of the electronic waves as the modulation of the current can be close to 100\%. It is important to stress the role of chirality in these experiments, as a way to decouple source and interferometer. Indeed, backscattering of electrons towards the source in non-chiral systems can lead to the modification of the source properties by the presence of the interferometer itself. An important difference between electrons and photons is also revealed in these experiments. Indeed, electrons interact with each others and this interaction tends to reduce the coherence of the electronic wavepacket which induces a reduction of the contrast \cite{Neder2006, Roulleau2009, Huynh} when varying the length difference between the interferometer arms.

The statistical properties of stationary sources have also been studied in the electronic analog of the Hanbury-Brown \& Twiss geometry \cite{Oliver1999,Henny1999,Henny1999a}. In this setup, a beam of electrons is partitioned on an electronic beamsplitter and the correlations $\langle I_t(t)I_r(t')\rangle$ between both transmitted $I_t(t)$ and reflected $I_r(t')$ intensities are recorded. The random partitioning on the beamsplitter is  a discrete process at the scale of individual particles: an electron (or a photon in optics) is either transmitted or reflected, so that the intensity correlations encode detailed information on the emission statistics of the source by comparing it with the reference of a poissonian process. In current experiments, the $t, t'$ time information is lost and the current fluctuations on long times are measured. For a dc biased ohmic contact, the regular and noiseless flow of electrons at the input of the splitter is reflected in the perfect anticorrelations of the output currents, $ \langle I_t  I_r\rangle = 0$.

The nature of the physical effects probed in these two types of experiments is quite different. Indeed, Mach-Zehnder interferometers probe the wave properties of the source, and interference patterns arise from a collection of many single-particle events. For light, classical analysis in terms of wave physics started during the 17th century (e.g. by Hooke, Huyghens)  to be further developed during the 18th and 19th centuries (e.g. by Young and Maxwell) and is associated with first order coherence function $\mathcal G^{(1)}(\mathbf{r},t;\mathbf{r'},t')=\langle E^{*}(\mathbf{r},t)E(\mathbf{r'},t')\rangle$, that encodes the coherence properties of the electric field $E(\mathbf{r},t)$ at position $\mathbf{r}$ and time $t$. The information obtained through Hanbury-Brown \& Twiss interferometry differ from a wave picture, as random partitioning on the beamsplitter is a discrete process, thus encoding information on the discrete nature of the involved particles. A classical model in terms of corpuscles can explain the features observed, and are described in optics using second order coherence function $\mathcal G^{(2)}(\mathbf{r},t;\mathbf{r'},t')=\langle E^{*}(\mathbf{r'},t')E^{*}(\mathbf{r},t)E(\mathbf{r},t)E(\mathbf{r'},t')\rangle$. The classical definitions of first and second order coherence of the electromagnetic field were extended by Glauber \cite{Glauber1963} to describe non-classical states of light by introducing the quantized electromagnetic field $\hat{E}(\mathbf{r},t)$. This description is currently the basic tool to characterize light sources in quantum optics experiments. It can be adapted to electrons in quantum conductors, and as in photon optics, both aspects of wave and particle nature of the carriers can be reconciled into a unified theory of coherence "à la Glauber".

Still, a few experiments cannot be understood within the wave nor the corpuscular description: this is the case when two-particle interferences effects related to the exchange between two indistinguishable particles take place.  The collision of two particles emitted at two different inputs of a beamsplitter can be used to measure their degree of indistinguishability. In the case of bosons, indistinguishable partners always exit in the same output. This results in a dip in the coincidence counts between two detectors placed at the output of the splitter when both photons arrive simultaneously on the splitter as observed by Hong-Ou-Mandel (HOM) \cite{Hong1987} in the late eighties. Fermionic statistics leads to the opposite behavior: particles exit in different outputs. This two particle interference effect has been observed using two stationary sources (dc biased contacts) and recording the reduction in the current fluctuations at the output of the splitter \cite{Liu1998}. The interference term could also be fully controlled \cite{Samuelsson2004, Neder2007} by varying the Aharonov-Bohm flux through a two-particle interferometer of geometry close to the Mach-Zehnder interferometer described above. In the latter case two-particle interferences can be used to post-select entangled electron pairs at the output of the interferometer. The production of a continuous flow of entangled electron-hole pairs has also been proposed using a beam splitter partitioning two edge channels \cite{Beenakker2003b}.

All these experiments emphasize the analogies between electron and photon propagation and provide important quantitative information on the electron source. They also show the differences between electron and photon optics, regarding the effect of Coulomb interaction or Fermi statistics. However, as particles are emitted continuously in the conductor, they miss the single particle resolution necessary to manipulate single particle states. In optics, the development of triggered single photon sources has enabled the manipulation and characterization of quantum states of light, opening the way towards the all-optical quantum computation \cite{Knill2001}. In electronics as well, several types of sources have been recently developed in quantum Hall edge channels, so that the field of electron quantum optics is now accessible.

In the first section, we introduce the formalism of electronic coherence functions as inspired by Glauber theory of light. It appears particularly suitable to describe the single electrons generated by triggered sources that we briefly review in the second section, focusing on the mesoscopic capacitor used as a single electron source. The use and study of short time current correlations to unveil the statistical properties of a triggered emitter are presented in the third section. We then discuss the two particle exchange interferences that take place in the Hanbury-Brown \& Twiss interferometer and analyze how these effects can be revealed in the partitioning of a single source as well as in a controlled two-electron collision. Finally the crucial issue of interactions between electrons and their impact on electron quantum optics experiments is discussed in the last section.

\section{Electron optics formalism}
\label{Formalism}
 A single edge channel is modeled as a one dimensional wire along which the electronic propagation is chiral, ballistic and spin polarized. The electronic degrees of freedom are described by the fermion field operator $\hat{\Psi}(x,t)$ that annihilates one electron at time $t$ and position $x$ of the edge channel, or equivalently, in the Fourier representation, by the operator $\hat{a}(\epsilon)$ that annihilates one electron of energy $\epsilon$ in the channel. Neglecting here Coulomb interactions which effects will be discussed in section \ref{Interactions}, the free propagation of the fermionic field simply corresponds to the forward propagation of electronic waves at constant velocity $v$:
\begin{equation}
\hat{\Psi}(x,t) = \frac{1}{\sqrt{h v}} \int d\epsilon \; \hat{a} (\epsilon) e^{i \frac{\epsilon}{\hbar} (x/v- t)} \label{Eqlibre}
\end{equation}
This time evolution is particularly simple as the fermion field operator $\hat{\Psi}(x,t)$ only depends on $x$ and $t$ through the difference $x-vt$. %As a consequence, we will leave the $x$ dependence and only retain the $t$ dependence.

\subsection{Electron-photon analogies}

The ballistic propagation of electrons along quantum Hall edge channels bears strong similarities with the propagation of photons in vacuum. These profound analogies can be noticed in the formalism describing the dynamics of the fermion field operator $\hat{\Psi}(x,t)$ on the one hand and the electric field operator, $\hat{E}(x,t) = \hat{E}^{+}(x,t) + \hat{E}^{-}(x,t)$ in quantum optics on the other hand \cite{Loudon1983}:
\begin{eqnarray}
\hat{E}^{+}(x,t)& = &  i \int d\epsilon  \sqrt{\frac{\epsilon}{2 h c \varepsilon_0 S}}\; \hat{a}(\epsilon) \; e^{i \frac{\epsilon}{\hbar}(x/c - t)} \label{E+}\\
\hat{E}^{-}(x,t)& = & \big(\hat{E}^{+}(x,t) \big)^{\dag}
\end{eqnarray}
Where $S$ is the transverse section perpendicular to the one dimensional propagation along the $x$ direction and $c$ the celerity of light propagation. For simplicity the polarization of the electric field has been omitted. From Eq.(\ref{E+}) one can see that the fermion field operator $\hat{\Psi}(x,t)$ is very similar to $\hat{E}^{+}(x,t)$, the part of the electric field that annihilates photons, where the complex conjugate $\hat{\Psi}^{\dag}(x,t)$ is similar to $\hat{E}^{-}(x,t)$ that creates photons. The electrical current $\hat I(x,t)$ in electron optics will then be the analog to the light intensity $\hat I_{ph}(x,t)$ in usual photon optics:
 \begin{equation}
\hat I(x,t) = -e v \hat{\Psi}^{\dag}(x,t) \hat{\Psi}(x,t) \quad \hat I_{ph}(x,t) = \hat{E}^{-}(x,t) \hat{E}^{+}(x,t)
 \end{equation}
More generally, the coherence properties of electron sources can be studied by characterizing the first order coherence $\mathcal{G}^{(1,e)}(x,t;x',t')$ \cite{Grenier2011a, Grenier2011NJP} defined in full analogy with Glauber's theory of optical coherences \cite{Glauber1963} with $\hat{\Psi}(x,t)$ replacing $\hat{E}^{+}(x,t)$. However, as $\hat{\Psi}(x,t)$ only depends on $x$ and $t$ through the difference $x -vt$, we will only retain the time dependence of $\mathcal{G}^{(1,e)}$ and set $x=x'=0$ in the rest of the manuscript:
\begin{eqnarray}
\mathcal{G}^{(1,e)}(t,t') & =& \langle \hat{\Psi}^{\dag}(t')\hat{\Psi}(t) \rangle
 \end{eqnarray}
 The first order coherence can also be defined for holes, $\mathcal{G}^{(1,h)}(t,t')  = \langle \hat{\Psi}(t')\hat{\Psi}^{\dag}(t) \rangle$ and is directly related to the electron coherence, $\mathcal{G}^{(1,h)}(t,t')= \frac{\delta(t-t')}{v} - \mathcal{G}^{(1,e)}(t,t')^{*}$. We will thus use mainly the electron coherence, the expression for holes will be used when it simplifies the notations.
The diagonal part, $t=t'$, of the first order coherence represent the 'populations' of the electronic source per unit of length, that is the electronic density which is proportional (with a factor $-ev$) to the electrical current at time $t$. The off-diagonal parts represent the coherences that are probed in an electronic interference experiment. In an equivalent way, coherence properties can also be defined in Fourier space:
\begin{eqnarray}
\tilde{\mathcal{G}}^{(1,e)}(\epsilon,\epsilon') & =& \frac{h}{v} \langle \hat{a}^{\dag}(\epsilon')\hat{a}(\epsilon) \rangle
 \end{eqnarray}
The diagonal elements, or populations, are then proportional to the number of electrons per unit energy while the off diagonal terms represent the coherences in energy space. It is worth noticing that in the case of a stationary emitter ($\mathcal{G}^{(1,e)}(t,t') = \mathcal{G}^{(1,e)}(t-t')$), these off diagonal terms in energy space vanish and the first order coherence can be characterized by the populations in energy only: $\tilde{\mathcal{G}}^{(1,e)}(\epsilon,\epsilon') \propto \delta(\epsilon-\epsilon')$.

\subsection{Electron-photon differences}
\label{ElectronPhotonDiff}
 Despite the deep analogies between electron and photon optics, some major differences remain. The first and most obvious one comes from the Coulomb interaction that affects electron and hole interactions. Contrary to photons, the propagation of a single elementary excitation is a complex many-body problem as one should consider its interaction with the large number of surrounding electrons that build the Fermi sea. This interaction leads in general to the relaxation and decoherence of single electronic excitations propagating in the conductor and will be discussed in section \ref{Interactions}. However, the free dynamics described by Eqs. (\ref{Eqlibre}) that neglects interaction effects already capture many interesting features of electronic propagation in ballistic conductors that will be discussed first. Another major difference is related to the statistics, fermions versus bosons, with important consequences on the nature of the vacuum. At equilibrium in a conductor, many electrons are present and occupy with unit probability states up to the Fermi energy $\epsilon_F$. The equilibrium state of the edge channel at temperature $T$ will be labeled as $| F \rangle$. As a first consequence, and contrary to optics, even at equilibrium, the first order coherence function does not vanish due to the non-zero contribution from  the Fermi sea which we label $\mathcal{G}^{(1,e)}_F(t,t') = \langle F  | \hat{\Psi}^{\dag}(t')\hat{\Psi}(t) | F \rangle $. It can be more easily computed in Fourier space, $\tilde{\mathcal{G}}^{(1,e)}_F(\epsilon,\epsilon')= \frac{h}{v} f(\epsilon) \delta(\epsilon-\epsilon')$ where it is diagonal and thus characterized by the population in each energy state given by the Fermi distribution $f(\epsilon)$ at temperature $T$.
 We will therefore consider the deviations of the first order coherence function compared to the equilibrium situation: $\Delta \mathcal{G}^{(1,e)}(t,t')  = \mathcal{G}^{(1,e)}(t,t') - \mathcal{G}^{(1,e)}_F(t,t')$. The electrical current carried by the edge channel does not vanish as well at equilibrium, $I_F= \langle F | \hat I(t) | F \rangle = -\frac{e}{h}  \int d\epsilon \; f(\epsilon) $. This equilibrium current is canceled by the opposite equilibrium current carried by the counterpropagating edge channel located on the opposite edge of the sample. In an experiment, the current is measured on an ohmic contact which collects the total current, difference between the incoming current carried by one edge and the outgoing current carried by the counterpropagating edge. The ohmic contact plays the role of a reservoir at thermal equilibrium such that the outgoing edge is at thermal equilibrium and carries the current $I_F$. The total current measured is then,  $:I(t): = I(t)-I_F$. In the following, in order to lighten the notations, $I(t)$ will refer to the total current, the Fermi sea contribution will always be subtracted, $I(t)=:I(t):$. The measurement of the electrical current on an ohmic contact thus characterizes the deviation of the state of a quantum Hall edge channel compared to its equilibrium state. It is proportional to the diagonal terms of the excess first order coherence of the source in time domain.
 \begin{equation}
  I(t)= -ev \Delta \mathcal{G}^{(1,e)}(t,t)
 \end{equation}
In term of elementary excitations, deviations from the Fermi sea consist in the creation of electrons above the Fermi sea and the destruction of electrons below it, or equivalently, the creation of holes of positive energy. Contrary to optics, where all the photons contribute with a positive sign to the light intensity, two kinds of particles with opposite charge and thus opposite contributions to the electrical current are present in electron optics. As we will see in the following of this manuscript, the propagation of carriers of opposite charge related to the presence of the Fermi sea leads to important differences with optics. Excess electron $\delta n_e(\epsilon)$ and hole $\delta n_h(\epsilon)$ populations are related to the diagonal terms, the populations, of the excess first order coherence in Fourier space:

 \begin{eqnarray}
 \delta n_e(\epsilon) & = & \frac{v}{h} \Delta \tilde{\mathcal{G}}^{(1,e)}(\epsilon, \epsilon) \\
 \langle \delta N_e \rangle & =& \int_{0}^{+\infty} d\epsilon \;\delta n_e(\epsilon) \\
 \delta n_h(\epsilon) & =& \frac{v}{h} \Delta \tilde{\mathcal{G}}^{(1,h)}(-\epsilon, -\epsilon)  = -\delta n_e(-\epsilon) \\
 \langle \delta N_h \rangle & =& \int_{0}^{+\infty} d\epsilon \;\delta n_h(\epsilon)
 \end{eqnarray}

\subsection{Stationary source versus single particle emission}
\label{StationarySES}
Stationary sources are the most commonly used in electron optics experiments and are implemented by applying a stationary bias $V$ to an ohmic contact which shifts the chemical potential of the edge channel by $-eV$. For such a stationary source, the first order coherence function does not depend separately on both times $t$ and $t'$  but only on the time difference $t-t'$. As already mentioned, such a source is fully characterized by its diagonal components in Fourier space  $\mathcal{G}^{(1,e)}(\epsilon,\epsilon') \propto \delta(\epsilon- \epsilon')$. In the case of the voltage biased ohmic contact, the electron population is simply given by the difference of the equilibrium Fermi distributions with and without the applied bias : $\Delta \tilde{\mathcal{G}}^{(1,e)}(\epsilon,\epsilon') = \frac{h}{v} \big[ f(\epsilon + eV) - f(\epsilon) \big] \delta(\epsilon-\epsilon')  $. The corresponding total number of electrons emitted per unit of energy in a long but finite measurement time $T_{meas}$ is then given by $\delta n_e(\epsilon) = \big[ f(\epsilon + eV) - f(\epsilon) \big] \frac{T_{meas}}{h}$. As mentioned in section \ref{elecopticsexp}, many electron optics experiments have been performed with this source to investigate the coherence properties of these sources using electronic interferometers.

\begin{figure}[h!]
\centering\includegraphics[width=\columnwidth]{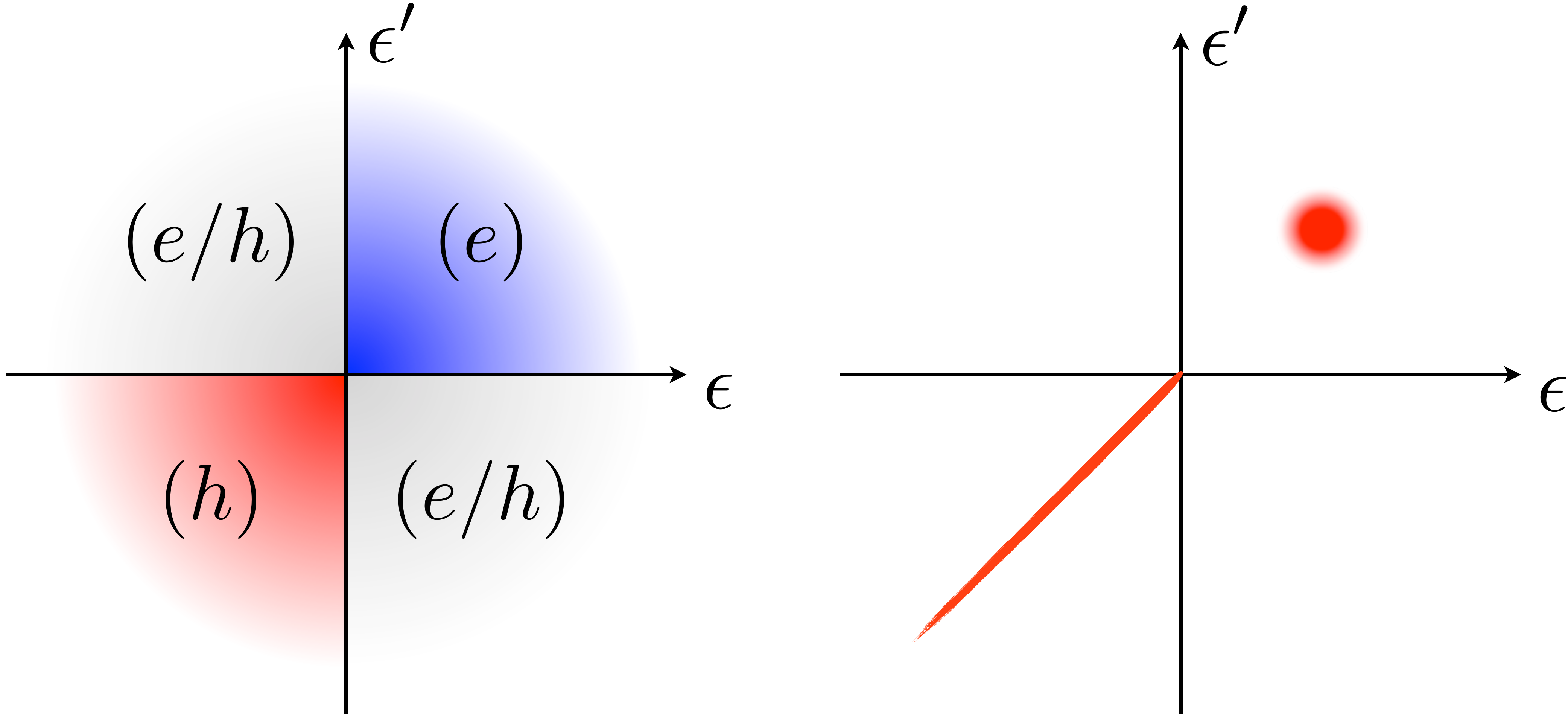}
\caption{Left panel : Quadrants of the electronic coherence function in Fourier space : electron $(e)$, hole $(h)$, and mixed electron/hole $(e/h)$. Right panel : schematic representation of a single-electron state created on top of the Fermi sea : the Fermi sea is represented by the half-diagonal $\epsilon=\epsilon'<0$ with no transverse extension. The single-electron state is pictured by a dot in the $(e)$-quadrant.}\label{fig:Quadrants}
\end{figure}

A different route of electron optics is the study of the propagation and the manipulation of single particle (electron or hole) states. Such a single electron state corresponding to the creation of one additional electron in wave function $\phi^e(x)$ above the Fermi sea can be formally written as:
\begin{eqnarray}
 \hat{\Psi}^{\dag} [\phi^e]|F \rangle & =&  \int dx \; \phi^{e}(x) \hat{\Psi}^{\dag}(x) |F \rangle
 \end{eqnarray}
 where $\phi^{e}(x)$ is the electronic wave function which Fourier components $\tilde{\phi}^{e}(\epsilon)$ are only non-zero for $\epsilon>0$,  corresponding to the filling of electronic states above the Fermi energy (at finite temperature, the single particle state has to be separated from the thermal excitations of the Fermi sea). This state is fully characterized by the first order coherence function:
 \begin{eqnarray}
 \Delta \mathcal{G}^{(1,e)}(t,t')& =&  \phi^{e}(-vt) \phi^{e,*}(-vt') \\
 \Delta \tilde{\mathcal{G}}^{(1,e)}(\epsilon,\epsilon')& =&  \tilde{\phi}^{e}(\epsilon) \tilde{\phi}^{e,*}(\epsilon') \label{GSES}
 \end{eqnarray}
 In a two dimensional ($\epsilon, \epsilon'$) representation of the first order coherence in Fourier space, such a single electron state can be represented as a spot in the $\epsilon>0, \epsilon'>0$ quadrant (see Fig.\ref{fig:Quadrants}, right panel). This quadrant thus corresponds to the electron states. The coherence of the wave function appears in the off diagonal components ($\epsilon \neq \epsilon'$) which clearly enunciates the fact that such single particle states cannot be generated by a stationary emitter but requires the use of a triggered ac source. These single electron emitters open a new route for electronic transport, where the object of study is an electronic wavefunction that evolves in time instead of the set of occupation probabilities for the electronic states. The study of such a source and its ability to produce single electron states will be the purpose of the next section.

 Note that the symmetric situation of a single hole creation can be described by the following state: $\hat{\Psi} [\phi^h]|F \rangle= \int dx  \; \phi^{h,*}(x) \hat{\Psi} (x) |F \rangle$ where  $\tilde{\phi}^{h}(\epsilon)$ has only non vanishing components for $\epsilon<0$ corresponding to electronic states below the Fermi energy. Its first order coherence function, $\Delta \tilde{\mathcal{G}}^{(1,e)}(\epsilon,\epsilon') = - \tilde{\phi}^{h}(\epsilon) \tilde{\phi}^{h,*}(\epsilon') $ corresponds to a spot in the $(\epsilon<0,\epsilon'<0)$-quadrant of hole states (see Fig.\ref{fig:Quadrants}). Note that the minus sign reflects the fact that a hole is an absence of electron in the Fermi sea.

 The two remaining quadrants ($\epsilon >0, \epsilon' <0$ and $\epsilon <0, \epsilon'>0$) in the $(\epsilon,\epsilon')$-plane are called the electron/hole coherences. They can be understood as the manifestation of a non fixed number of excitations (electrons and holes) which characterizes states that are neither purely electron nor purely hole states. An example of such a state can be written as:
 \begin{equation}
 |\Psi \rangle =  \alpha |F \rangle + \beta \int dx  dx' \; \phi^{h^,*}(x) \phi^{e}(x') \hat{\Psi} (x) \hat{\Psi}^{\dag} (x')|F \rangle
 \end{equation}
This state is the coherent superposition of the equilibrium state and a non-equilibrium state that corresponds to the creation of one electron and one hole (one electron/hole pair). The total number of particles stays fixed but the number of excitations is not, such that this state cannot be seen as a pure 'electron-hole' pair. By computing the first order coherence function in Fourier space, one gets (zero temperature has been assumed for simplicity):
\begin{eqnarray}
\Delta \mathcal{G}^{(1,e)}(\epsilon,\epsilon') & =  & |\beta|^2 \big[\tilde{\phi}^{e}(\epsilon) \tilde{\phi}^{e,*}(\epsilon') -\tilde{\phi}^{h}(\epsilon) \tilde{\phi}^{h,*}(\epsilon')\big] \nonumber \\
& -&  \alpha^* \beta \tilde{\phi}^{e}(\epsilon) \tilde{\phi}^{h,*}(\epsilon') - \alpha \beta^* \tilde{\phi}^{h}(\epsilon) \tilde{\phi}^{e,*}(\epsilon') \nonumber \\
 \end{eqnarray}
The first two terms correspond to the electron and hole states discussed previously.  The last two terms correspond to spots in the electron/hole quadrants of the $(\epsilon,\epsilon')$-plane. This kind of terms will appear when the source fails to create a well defined number of electron/hole excitations but rather a coherent superposition of states with different number of excitations.

\section{Single electron emitters}
\label{SES}

\subsection{Generation of quantized currents}

The first manipulations of electrical currents at the single charge scale have been implemented in metallic electron boxes. In these systems, taking advantage of the quantization of the charge,  quantized currents could be generated in single electron pumps with a repetition frequency of a few tens of MHz \cite{Geerligs1990, Pothier1992}. These single electron pumps have been realized almost simultaneously in semiconducting nanostructures \cite{Kouwenhoven1991} where the operating frequency was recently extended to GHz frequencies \cite{Blumenthal2007,Maire2008}. These technologies have also been implemented under a strong magnetic field  \cite{Giblin2010,Leicht2011,Giblin2012}, to inject electrons in high-energy quantum Hall edge channels. Another route for quantized current generation is to trap a single electron in the electrostatic potential generated by a surface acoustic wave propagating \cite{Shilton1996,Talyanskii1997} through the sample. This technique has recently enabled the transfer of single charges between two distant quantum dots \cite{Hermelin2011, McNeil2011}. However, even if these devices are good candidates to generate and manipulate single electron quantum states in one dimensional conductors, their main applications concern metrology and a possible quantum representation of the ampere (for a review on single electron pumps and their metrological applications, see \cite{Pekola2012}).

Another proposal to generate single particle states in ballistic conductors, and which relies on a much simpler device, has been proposed \cite{Levitov1996,Ivanov1997,Keeling2006,Dubois2013}: the DC bias applied to an ohmic contact, and that generates a stationary current, is replaced by a pulsed time dependent excitation $V_{exc}(t)$. For an arbitrary time dependence and amplitude of the excitation, such a time dependent bias generates an arbitrary state that, in general, is not an eigenstate of the particle number but is the superposition of various numbers of electron and hole excitations. The differences of such a many body state compared to the creation of a single electronic state above the Fermi sea can be outlined using the first order coherence function of the source, $\Delta \mathcal{G}^{(1,e)}(\epsilon,\epsilon')$ in Fourier space. Contrary to the single electronic excitation which has only non-zero values in the electron domain $\epsilon,\epsilon'>0$, such a state has also non zero values in the hole sector ($\epsilon,\epsilon'<0$) representing the spurious hole excitations generated by the source. Finally, in this case, $\Delta \mathcal{G}^{(1,e)}(\epsilon,\epsilon')$ also exhibits non zero electron-hole coherences as such a state is not an eigenstate of the excitation number. It can be shown that by applying a specific Lorentzian shaped pulse containing a quantized number of charges: $e^2/h \; \int dt V(t) = n e$, exactly $n$ electronic excitations could be generated in the electron sector without creating any hole excitation. In particular, the voltage $V(t)=\frac{h \tau_0/\pi}{t^2 + \tau_0^2}$ generates a single electron above the Fermi sea as recently experimentally demonstrated \cite{Dubois2013b}.

We followed a different route to generate single particle states which bears more resemblance with the single electron pumps mentioned before. The emitter, called a mesoscopic capacitor consists in a quantum dot capacitively coupled to a metallic top gate and tunnel coupled to the conductor. Compared to the pumps presented above, only one tunnel barrier is necessary such that the device is easier to tune. This difference implies that the source is ac driven and thus generates a quantized ac current whereas pumps generate a quantized dc current (note that in a recent proposal, Battista et al. \cite{Battista2011,Battista2012} suggested a new geometry where the electron and hole streams are separated, such that a dc current is generated). Compared to Lorentzian pulses, the single particle emission process does not depend much on the exact shape of the excitation drive. The quantization of the emitted current is ensured by the charge quantization in the dot. Another difference comes from the possibility to tune the energy of the emitted particle, as emission comes from a single energy level of the dot which energy can be tuned to some extent.

\subsection{The mesoscopic capacitor}
\label{CapaMeso}
\begin{figure}[h!]
\centering\includegraphics[width=\columnwidth]{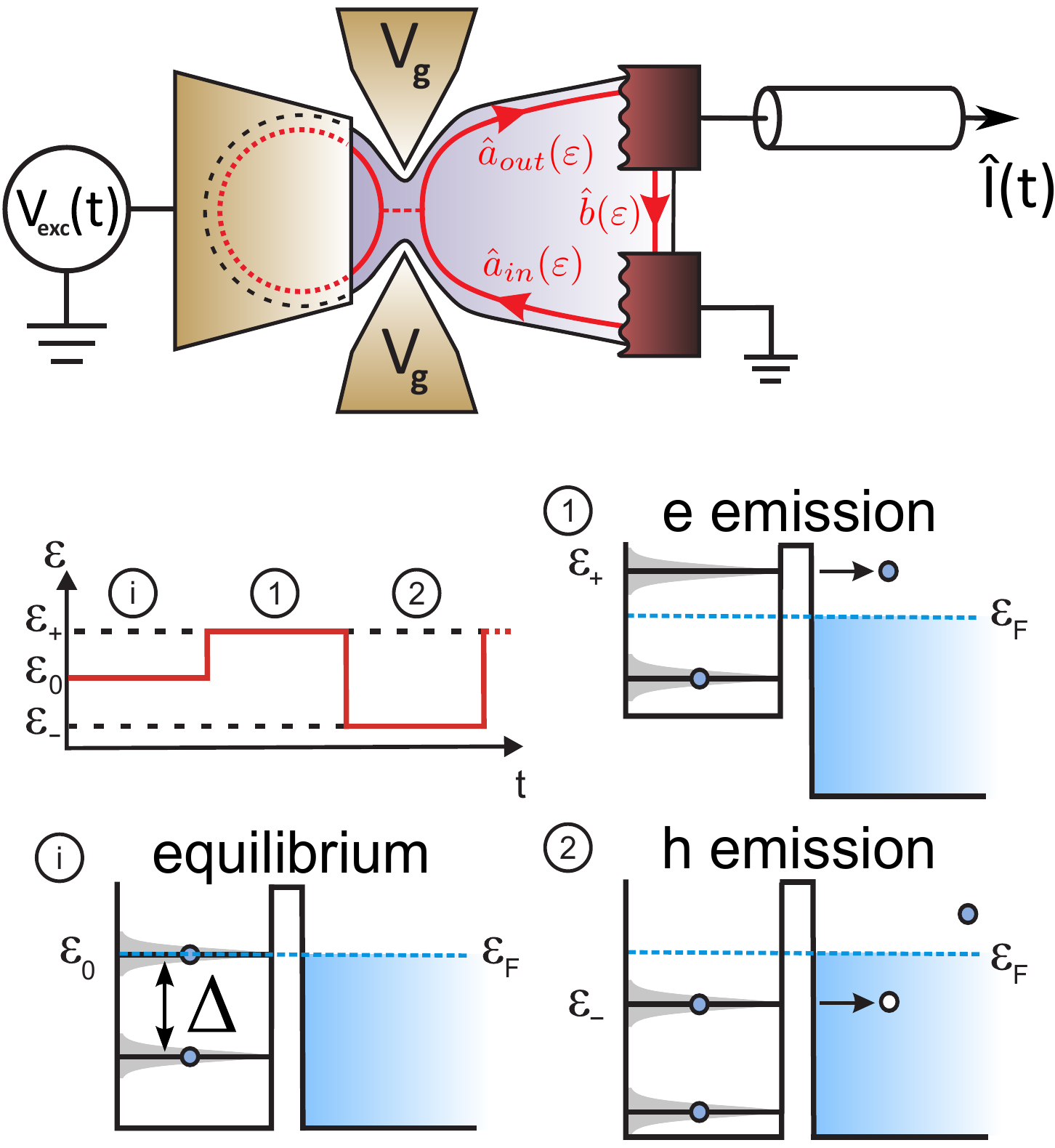}
\caption{The mesoscopic capacitor. Upper panel, sketch of the mesoscopic capacitor. Lower panel, sketch of single electron/hole emission process.
\label{fig2_1} }
\end{figure}

The mesoscopic capacitor \cite{Buttiker1993a, Gabelli2006a, Parmentier2012} is depicted in Fig.\ref{fig2_1}. It consists
of a submicron-sized cavity (or quantum dot) tunnel coupled
to a two-dimensional electron gas through a quantum point contact
(QPC) whose transparency $D$ is controlled by the gate
voltage $V_\mathrm{g}$. The potential of the dot is controlled by a metallic top gate deposited on top of the dot and capacitively coupled to it. This conductor realizes the
 quantum version of a RC circuit, where the dot and electrode
define the two plates of a capacitor while the quantum point
contact plays the role of the resistor.  As mentioned in the first section, a large perpendicular
magnetic field is applied to the sample in order to reach the
integer quantum Hall regime, and we consider the
situation where a single edge channel is coupled to the dot.
Electronic transport can thus be described by the propagation of
spinless electronic waves in a one dimensional conductor.
Electrons in the incoming edge channel can tunnel onto the quantum
dot with the amplitude $\sqrt{\mathcal{D}}=\sqrt{1-r^2}$, perform
several round-trips inside the cavity, each taking the finite time
$\tau_0 =l/v$ ($l$ is the dot circumference), before finally tunneling back out into the
outgoing edge state. In these expressions, the reflection
amplitude $r$ has for convenience been assumed to be real and
energy-independent. For a micron
size cavity, $\tau_0$ typically equals a few tens of picoseconds. As a result of these coherent oscillations inside the electronic cavity, the propagation in the quantum dot can be described by a discrete
energy spectrum with energy levels that are separated
by a constant level spacing $\Delta$ related to the time of one round-trip $\Delta=h/\tau_0$, see Fig. 3. The levels
are broadened by the finite coupling between the quantum
dot and the electron gas, determined by the QPC
transmission $D$. This discrete spectrum can be shifted compared to the Fermi energy first in a static manner, when a static potential $V_0$ is applied to the top gate, but also dynamically, when a time dependent excitation $V_{exc}(t)$ is applied.
When a square shape excitation is applied, it
causes a sudden shift of the quantum dot energy
spectrum. We consider the optimal situation where the highest occupied energy level is initially located at energy $\epsilon_F$  at resonance with the Fermi energy in the absence of drive (labelled by $i$ on Fig.\ref{fig2_1}, lower panel).
When a square drive is applied with a peak to peak amplitude $2 e V_{exc}$ comparable to $\Delta$, an electron is emitted above the Fermi energy from the highest occupied energy level in the first half period (labeled as 1 in
Fig.\ref{fig2_1}), an electron is then absorbed from the
electron gas (corresponding to the emission of a hole as
indicated in Fig.\ref{fig2_1}) in the second half period (labeled as 2 in
Fig.\ref{fig2_1}). Repeating this sequence at a
drive frequency of $f\sim 1$ GHz thus gives rise to periodic
emission of a single electron followed by a single hole \cite{Feve2007a}.
Previous discussion neglects the effects of Coulomb interaction inside the dot.
It is characterized by the charging energy $E_c = \frac{e^2}{C_g}$, where $C_g$ is the geometrical capacitance of the dot.
It adds to the orbital level spacing $\Delta$ in the addition energy of the dot $\Delta^{*} = \Delta + \frac{e^2}{C_g}$ that defines the energetic cost associated with the addition or removal of one electron in the dot. It is thus the relevant energy scale for charge transfers between the dot and the edge channel. However, the magnitude of Coulomb interaction effects has been estimated to be of the same order as the orbital level spacing \cite{Parmentier2010}. This rather low contribution of interactions explains the success of the non-interacting models used throughout this manuscript to describe the dot. In these non-interacting models, we take the level spacing $\Delta$ to be equal to $\Delta^{*}$ which captures both orbital and interaction effects.
\medskip

This emission of a quantized number of particles by the dot can be first characterized through the current generated by the emitter averaged on a large number of emission sequences.

\subsection{Average current quantization}

\begin{figure}[h!]
\centering\includegraphics[width=\columnwidth]{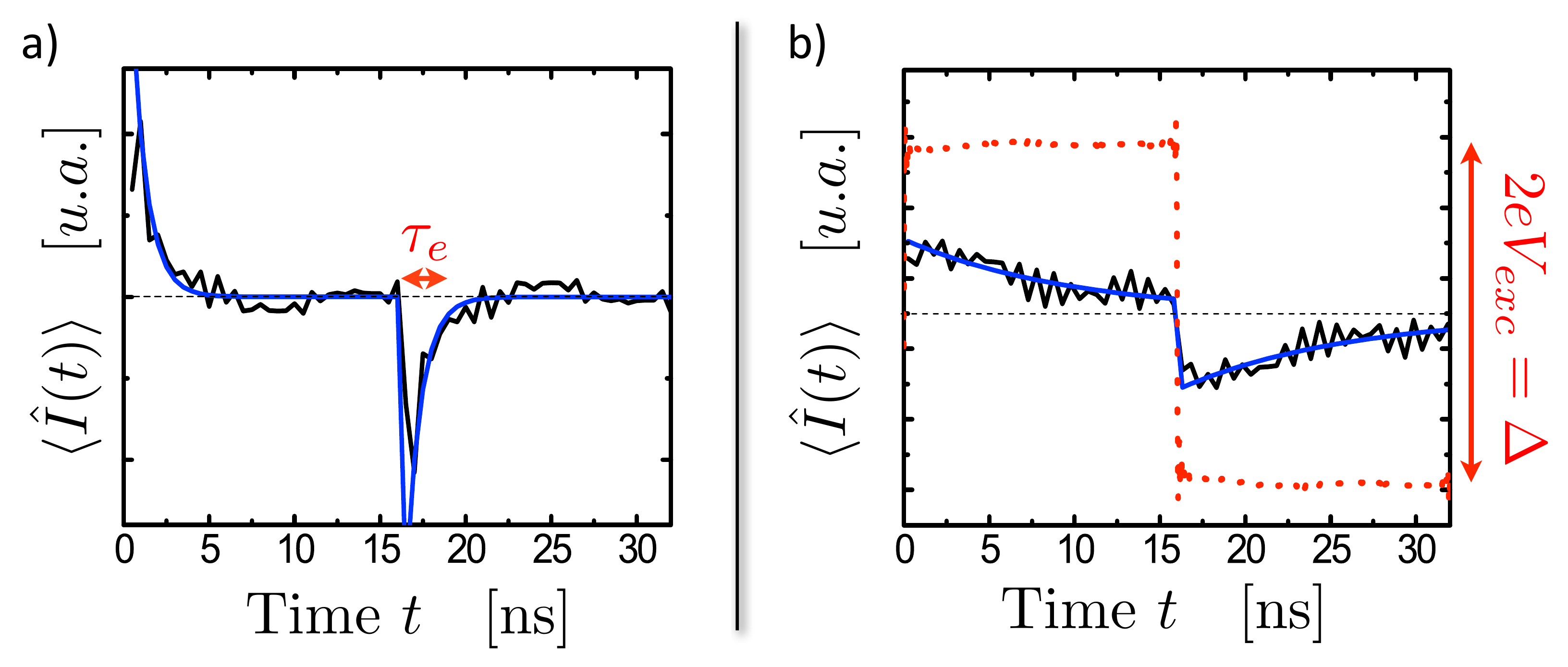}
\caption{Measurements of the average current in the time domain. The black traces represent the experimental points while the blue trace is an exponential fit. The escape times and average transmitted charges are $\tau_e = 0.9$ ns, $\langle Q_t \rangle = e$, left panel, $\tau_e = 10$ ns, $\langle Q_t \rangle = 0.7 e$, right panel. The red dotted line represents the square excitation voltage.
}\label{fig2_2}
\end{figure}

An important characteristic of the mesoscopic capacitor lies in its capacitive coupling, such that it cannot generate any dc current. This emitter is intrinsically an ac emitter and, as such, can be characterized through ac measurements of the current averaged on a large number of electron/hole emission cycles. This current $\langle I(t) \rangle $ can first be measured in time domain \cite{Mahe2008},  using a fast averaging card with a sampling time of $500$ ps and averaging on approximately $10^8$ single electron/hole emission sequences. To get a good resolution on the time dependence of the current, this card limits the drive frequency to a few tens of MHz. The resulting current generated by the source for a drive frequency of 32 MHz is represented on Figure \ref{fig2_2}. We observe an exponential decay of the current with a positive contribution that corresponds to the emission of the electron followed by its opposite counterpart that corresponds to the emission of the hole. This exponential decay corresponds to what one would naively expect for a RC circuit. At $t=0$ the square excitation triggers the charge emission by promoting an occupied discrete level above the Fermi energy which is then coupled to the continuum of empty states in the edge channel. The probability of charge emission, and hence the current, follows an exponential decay on an average time governed by the transmission $D$ and the level spacing, $\tau_e = h/D \Delta$ \cite{Moskalets2013a}. On Fig. \ref{fig2_2} (left panel), the escape time is $\tau_e=0.9$ ns, much smaller than the half period, such that the electron is allowed enough time to escape the dot. This is reflected by the measured quantization of the average transmitted charge \cite{Feve2007a,Mahe2008}, $\langle Q_t \rangle = \int_0 ^{T_0/2} \langle I(t) \rangle dt =e$ (where $T_0=1/f=2\pi/\Omega$ is the period of the excitation drive), which shows that one electron and hole are emitted on average by the source. By tuning the transmission, the escape time can be controlled and varied. On Fig. \ref{fig2_2} (right panel) $\tau=10$ ns which is comparable with the half-period. In this situation, some single electron events are lost and the average emitted charge is not quantized anymore, which defines a probability of charge emission $P$, $\langle Q_t \rangle =P e <e$ ($P=0.7$ for $\tau_e=10$ ns ). For an exponential decay of the current, the emission probability can be easily computed, $P=\tanh(\frac{T_0}{4 \tau_e})$.

\begin{figure}
 \centering \includegraphics[width=\columnwidth]{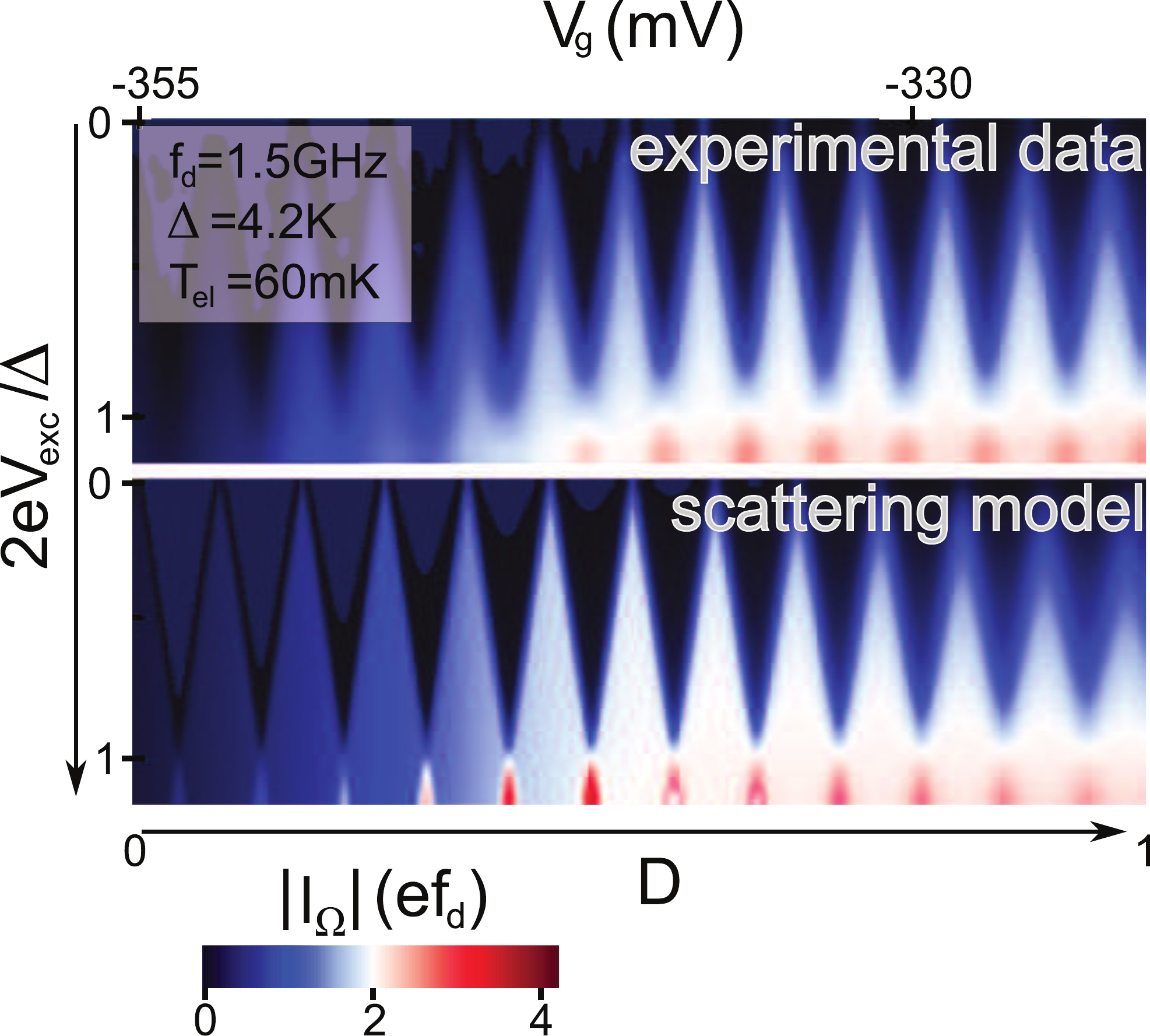}
   \caption{Two dimensional color plot of the modulus of the average current. The top figure represents the experimental points while the bottom figure is a simulation using Floquet scattering theory.  \label{fig2_3}}
\end{figure}

At higher frequencies (GHz frequencies), the dot cannot be characterized by current measurements in the time domain anymore as the limited $500$ ps resolution becomes larger than the half-period. In that case, we measure the first harmonic of the current $I_{\Omega}$ in modulus and phase using a homodyne detection. The quantization of the emitted charge is then reflected in a quantization of the current modulus $| I_{\Omega} | =2 e f$ while the escape time can be deduced from the measurement of the phase $\phi$, $\tan{ \phi} = \Omega \tau_e$. Fig. \ref{fig2_3} (upper panel) presents the measurement of the modulus of the current as a function of the dot transmission (horizontal axis) and the excitation drive amplitude (vertical axis). The value of the current modulus is encoded in a color scale. White diamonds correspond to areas of quantized modulus of the ac current  $| I_{\Omega} | =2 e f$ \cite{Feve2007a}. These diamonds are blurred at high transmissions, where the charge quantization on the dot is lost due to charge fluctuations, they also vanish at small transmission when the average emission time becomes comparable or longer than the half period. This quantization of the average ac current is the counterpart, in the frequency domain, of the charge quantization for time domain measurements.

The single electron emitter can be very conveniently described by the scattering theory of electronic waves submitted to a time-dependent scatterer. As the scatterer is periodically driven, one can apply the Floquet scattering theory \cite{Moskalets2002, Moskalets2007, Moskalets2008, Parmentier2012}. Any physical quantity can be numerically computed from the calculation of the Floquet scattering matrix. In particular, Floquet calculations can be compared with the current modulus measurements plotted on Figure \ref{fig2_3} (simulations are on the lower panel), for any excitation drive $V_{exc}(t)$. The excitation is a square drive the electronic temperature is $T_\mathrm{el}\approx 60$ mK and the level spacing of the dot is  $\Delta=4.2$ K. The QPC gate voltage $V_g$ controls both the transmission $D(V_g)$ and the dot potential $V_0(V_g)$. For the transmission $D(V_g)$, we use a saddle-point transmission law \cite{Buttiker1990} with two parameters, for the potential $V_0(V_g)$, we use a capacitive coupling of the dot potential to the QPC gate characterized by a linear variation. Using these parameters, the agreement between the
experimental data and numerical calculations is very good, up to
small energy-dependent variations in the QPC transmission which
were not included in the model.

\section{Second order correlations of a single electron emitter} \label{chap3}
\subsection{Second order coherence function}
\begin{figure}[h!]
\centering\includegraphics[width=\columnwidth]{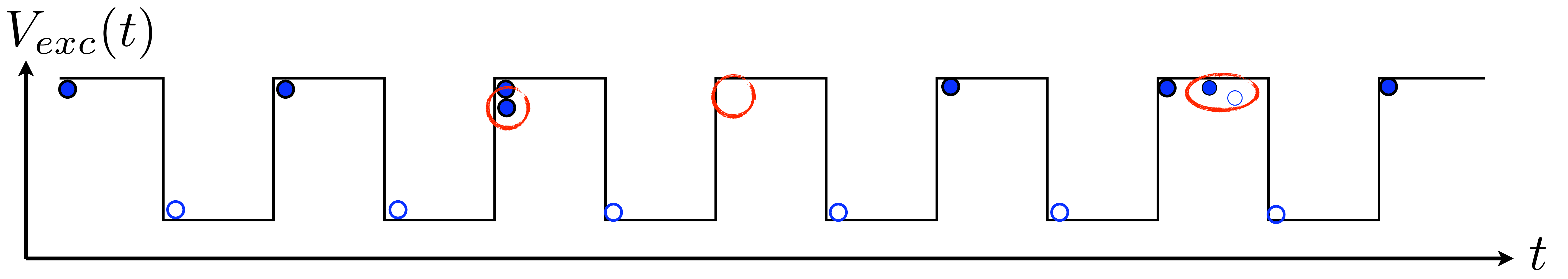}
\caption{\label{fig3_1} Sketch of electron/hole emission sequences. Electrons/holes are represented by blue/white dots. Spurious events are emphasized by red circles.}
\end{figure}

Although the measurement of the quantization of the charge emitted on one period is a strong indication that the source acts as an on-demand single particle emitter, it cannot be used as a demonstration that single particle emission is achieved at each of the source's cycles. The emitted charge is averaged on a huge number of emission periods, and hence does not provide any information on the statistics of electron emission. As can be seen on Fig.\ref{fig3_1}, the absence of electron emission on one cycle could be compensated by the emission of two electrons on the second one. An additional electron/hole pair could also be emitted in one cycle \cite{Keeling2008, Vanevic2008}. These various processes would not affect the average emitted charge and the quantization of the average current. In optics, single particle emission by photonic sources is demonstrated by the use of light intensity correlations \cite{Loudon1983,Michler2000,Yuan2002,Bozyigit2010}. In electronics as well, to demonstrate that exactly a single particle is emitted, one needs to go beyond the measurement of average quantities and study the correlations of the emitted current. Single particle emission can be demonstrated through the measurement of second order correlations functions of the electrical current.

\begin{figure}
\centering \includegraphics[width=\columnwidth]{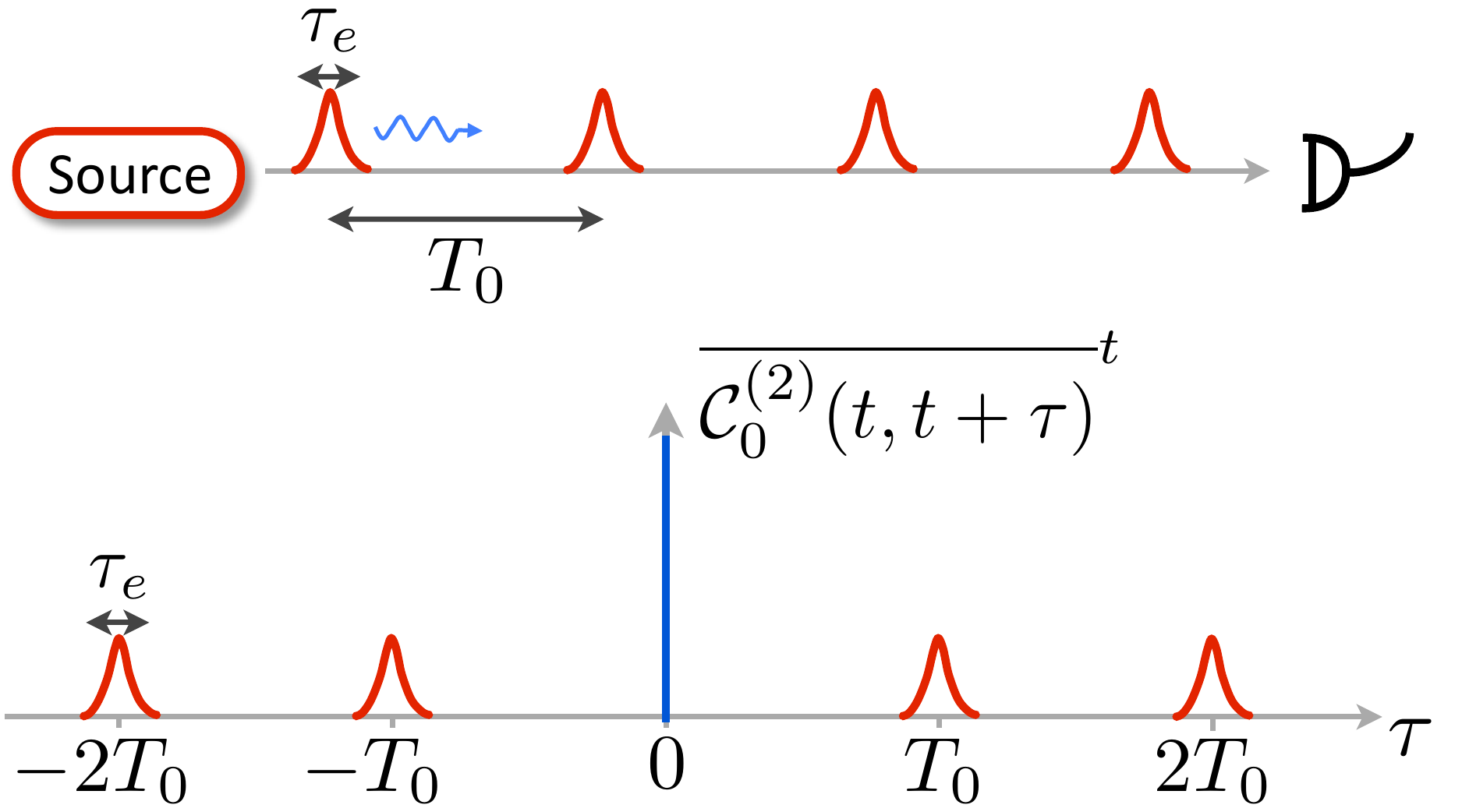}
   \caption{\label{fig3_6} Sketch of the second order correlation. Single particle wavepackets of width $\tau_e$ are emitted with period $T_0$. The blue trace represents the first term in Eq.(\ref{EqG0}) while the red trace represents the second term. The latter goes to zero on short time, reflecting that the source emits particles one by one. }
   \end{figure}

The second order correlation is usually defined by the joint probability to detect one particle at time $t$ and one particle at time $t'$. It reveals the correlations between particles, that is, their tendency to arrive close to each others (called bunching),  or on the opposite to be well separated (antibunching). Here, as we rely on current, or density measurements, we focus on the density-density correlation function. Using the fermion field operator at times $t$ and $t'$, it goes like:
\begin{eqnarray}
&&\mathcal{C}^{(2)}_0(t,t')  =  \langle \hat{\Psi}^{\dag}(t)\hat{\Psi}(t)\hat{\Psi}^{\dag}(t')\hat{\Psi}(t') \rangle \\
& & = \frac{\delta(t-t')}{v} \langle \hat{\Psi}^{\dag}(t)\hat{\Psi}(t) \rangle + \langle \hat{\Psi}^{\dag}(t')\hat{\Psi}^{\dag}(t)\hat{\Psi}(t) \hat{\Psi}(t') \rangle
\end{eqnarray}
The first term merely represents the autocorrelation of the charge at equal times and is proportional to the number of particles, that is to the average density. It is usually referred to as the shot noise term and reflects charge granularity. The second term is the joint probability to detect one particle at time $t$ and one particle at time $t'$ and encodes the correlations between particles. It is called the second order coherence function $\mathcal{G}^{(2)}(t,t')$ in a description "à la Glauber" of the electromagnetic field. In particular, if a single particle is present in the system (and only in this situation), this term vanishes for all times $t$, $t'$. It is therefore through the measurement of this term that single particle emission is asserted (in optics for example). Note that in many cases, and in particular, in the cases considered in this manuscript, the second order correlations can be expressed as a function of the first order ones through the use of Wick's theorem \footnote{In particular, Wick theorem can be applied to single particle states resulting from the addition of one electron or one hole or to the case of a periodically driven scatterer which we treat through Floquet scattering formalism. However, Wick theorem would not apply in the case where electron-electron interactions are present.}
\begin{eqnarray}
&&\mathcal{C}^{(2)}_0(t,t')  =   \frac{\delta(t-t')}{v} \mathcal{G}^{(1,e)}(t,t) +   \nonumber \\
&&  \mathcal{G}^{(1,e)}(t,t)\mathcal{G}^{(1,e)}(t',t') \Big[1 - \frac{|\mathcal{G}^{(1,e)}(t,t')|^2}{\mathcal{G}^{(1,e)}(t,t)\mathcal{G}^{(1,e)}(t',t')} \Big]
\label{EqG0}
\end{eqnarray}
Focusing the discussion on the second term which encodes the correlations between particles, we observe that perfect antibunching is always observed for $t=t'$ as two fermions cannot be detected at the same time due to Pauli exclusion principle. However, in general, two fermions can be detected at arbitrary times $t \neq t'$ except for a single particle state where the second term vanishes for arbitrary times $t,t'$. Indeed, ignoring first the presence of the Fermi sea, the single electron coherence of a single particle state reads, $\mathcal{G}^{(1)}(t,t') = \phi^{e}(-vt) \phi^{e,*}(-vt')$ such that $|\mathcal{G}^{(1)}(t,t')|^2 = \mathcal{G}^{(1)}(t,t)\mathcal{G}^{(1)}(t',t')$ for all times $t,t'$. In an experimental situation, the emission of a single particle state is periodically triggered with a period $T_0$. Considering an emitter with an average emission time $\tau_e$, the expected typical resulting trace for $\mathcal{C}^{(2)}_0(t,t')$ (averaged on the absolute time $t$) can be plotted on Fig.\ref{fig3_6}. The first term  in Eq.(\ref{EqG0}) is a Dirac peak and is plotted in blue. The second term is represented in red, lateral peaks centered on $t'-t = n \times T_0$ and of width $\tau_e$ correspond to the detection of two subsequent emission events separated by time $n T_0$. These peaks disappear on short times ($n=0$) as two different particles cannot be detected within the same emission period. This suppression is the hallmark of a single particle state: whenever two or more particles are emitted on the same emission period, this central peak would reappear.

However, one must be careful in the use of these arguments, as true single particle states are not available in quantum conductors due to the presence of the Fermi sea. We can only produce single particle states defined by the addition of one electron (or one hole) above (or below) the Fermi sea  which consists in a large number of electrons. It is thus not clear whether we can apply the above reasoning and use the second order correlation functions to detect states that result from the addition of a single electron above the Fermi sea (or equivalently the addition of a single hole below). We would also like to slightly change the definition of the second order correlation function in such a way that it can be directly expressed as a function of the natural observable of this system, that is, the electrical current. We thus adopt this new definition of the second order correlation function which is defined through the measurement of the excess current correlations at times $t$ and $t'$:
\begin{eqnarray}
\mathcal{C}^{(2)}(t,t') & = &  \langle \hat{I}(t')\hat{I}(t) \rangle - \langle \hat{I}(t')\hat{I}(t) \rangle_F \label{eq:g2a}
\end{eqnarray}
As seen in section \ref{ElectronPhotonDiff}, the second term is necessary to suppress the current correlations that already exist at equilibrium when the source is off. To enlighten the analogies between this expression and the previous definition that was valid in the absence of the Fermi sea, let us consider as previously a case where Wick theorem applies:
\begin{eqnarray}
\mathcal{C}^{(2)}(t,t')  & = &    \delta(t-t') \langle \hat{I}(t) \rangle  \nonumber \\
& + & \langle \hat{I}(t') \rangle \langle \hat{I}(t) \rangle  \Big[1 - \frac{|\Delta \mathcal{G}^{(1,e)}(t,t')|^2 }{\Delta \mathcal{G}^{(1,e)}(t,t) \Delta \mathcal{G}^{(1,e)}(t',t')}\Big]  \nonumber \\
& - &   e^2 v^2 \mathcal{G}^{(1,e)}_{F}(t,t') \; \Delta \mathcal{G}^{(1,e)}(t',t) \nonumber \\
& - & e^2 v^2 \mathcal{G}^{(1,e)}_{F}(t',t) \; \Delta \mathcal{G}^{(1,e)}(t,t')  \label{eq:g2b}
\end{eqnarray}

This expression presents many analogies with Eq.(\ref{EqG0}), in particular, the first two terms are identical except for the replacement of $\mathcal{G}^{(1,e)}$ by the contribution of the source only, $\Delta \mathcal{G}^{(1,e)}$.  These two terms thus provide a way to identify the single particle states generated by the source. However, the last two terms are not present in Eq.(\ref{EqG0}) as they represent correlations between the Fermi sea and the single particle source. Contrary to the first order correlation where the source and Fermi sea contributions could be separated, this is not the case in the second order correlations.

\subsection{High frequency noise of a single particle emitter}

\begin{figure}
\centering \includegraphics[width=\columnwidth]{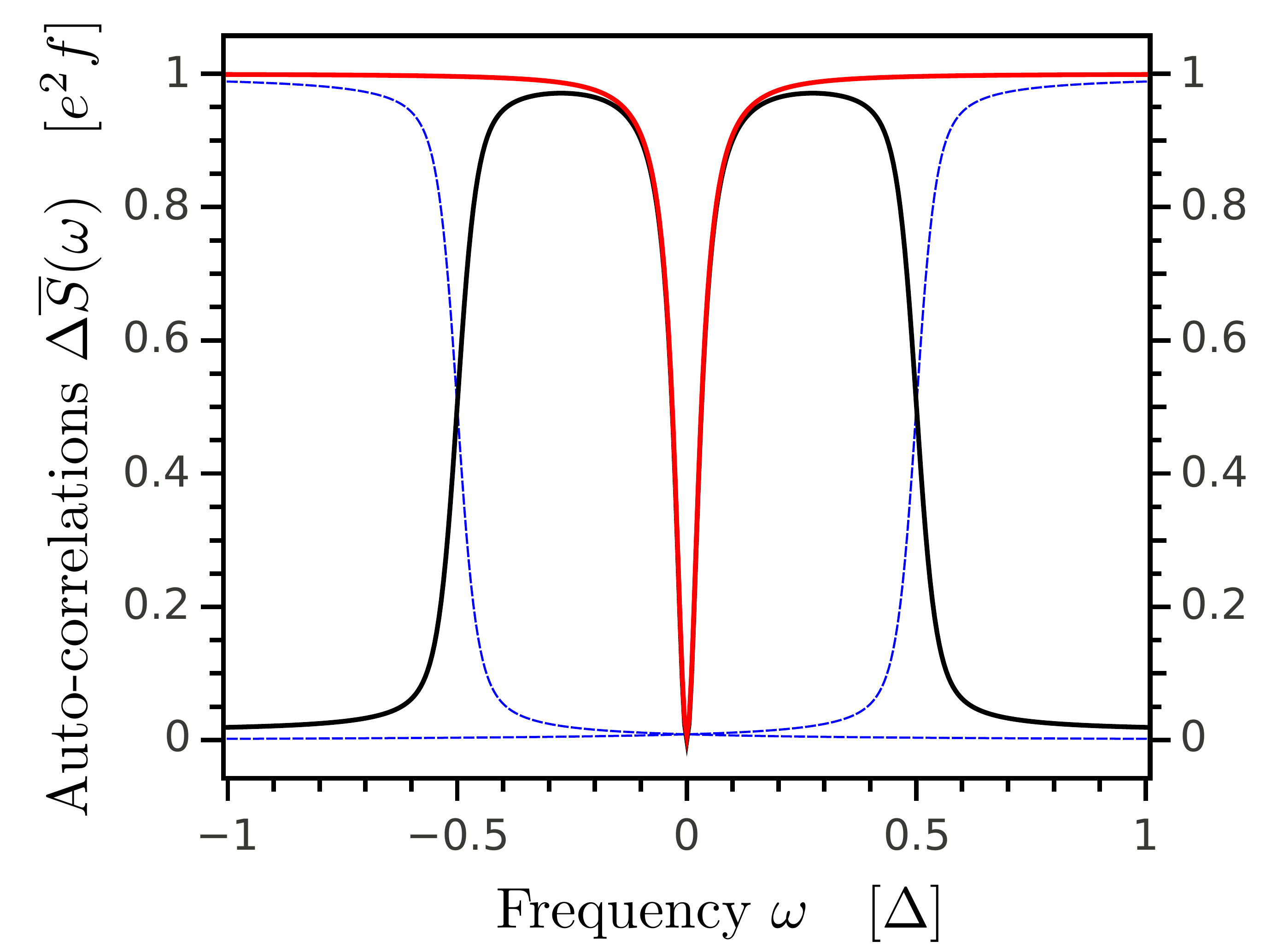}
   \caption{\label{fig3_2} Different terms of the noise spectrum $S(\omega)$ of a single particle emitter. The blue dashed line represents the Fermi sea contribution responsible for the high frequency cut-off, $-S_{F}$ defined from Eq.(\ref{EqSFS}), while the red trace is the noise spectrum neglecting the Fermi sea contribution. The black trace is the total noise spectrum obtained from the substraction of the blue dashed line to the red trace. }
\end{figure}

In electronics, current correlations are measured through the current noise spectrum $S(\omega)$. It is usually defined for a stationary process. For a non stationary process, it can be defined in analogy by performing an average on the current fluctuations on the absolute time $t$:
\begin{eqnarray}
S(\omega) & = &  2 \int d\tau \overline{\langle \delta I(t+\tau) \delta I(t) \rangle}^t e^{-i\omega \tau}
\end{eqnarray}
In the following, equilibrium noise contribution that can be measured when the source is off will always be subtracted from the noise spectrum in order to analyze the source contribution to the noise only. $S(\omega)$ is then directly given by the Fourier transform of the second order correlation defined above by Eqs. (\ref{eq:g2a}) and(\ref{eq:g2b}) up to an additional contribution related to the average current:
\begin{eqnarray}
S(\omega) & = &  2 \int d\tau \big[ \overline{\mathcal{C}^{(2)}(t,t+\tau) \rangle}^t + \overline{\langle \hat{I}(t+\tau) \rangle \langle \hat{I}(t) \rangle}^t \big]  e^{-i\omega \tau} \nonumber \\
\end{eqnarray}
The current noise spectrum provides a direct access to the second order correlation function and is thus an appropriate tool to demonstrate single particle emission. However, it is important to characterize the contribution to the noise spectrum of the last terms of Eq. (\ref{eq:g2b}), which we label $S_F(\omega)$ as these terms did not provide information on the source only but on correlations between the source and the Fermi sea.
\begin{eqnarray}
S_F(\omega) & =& - \frac{2 e^2}{T_{meas}}\int d\epsilon f(\epsilon)  \Big[ \delta n_{e}(\epsilon-\hbar \omega) +\delta n_e(\epsilon + \hbar \omega) \Big]  \nonumber \\ \label{EqSFS}
\end{eqnarray}
 To evaluate this contribution, let us consider a source that emits one electron at energy $\epsilon_e \approx \Delta/2$ above the Fermi sea. As $\delta n_{e}(\epsilon \pm \hbar \omega)$ represents the population of excitations emitted by the source at energy $\epsilon \pm \hbar \omega$, it is non zero when the energy is of the order of $\epsilon_e$. However, from the Fermi sea contribution $f(\epsilon)$, we have $\epsilon \leq 0$ which means that $S_F(\omega)$ becomes only non negligible at high frequency $ \hbar \omega \approx \epsilon_e$. Generally, this contribution can be safely neglected if the frequency is much lower than the energy of the excitations emitted by the source. Practically, this approximation holds for a measurement frequency $f \ll \frac{\Delta}{2 h}$, with $\frac{\Delta}{2 h}\gtrsim 20$ GHz. In the following, measurements were performed at $f \simeq 1$ GHz such that correlations between the source and the Fermi sea can safely be neglected in the noise measurements and the current correlations can be used to analyze the statistics of the source exactly as if the source was emitting in vacuum. From Eq. (\ref{eq:g2b}) we then directly obtain for a single particle emitter:
\begin{eqnarray}
 \mathcal{C}^{(2)}(t,t') &  = &  \delta(t-t') \langle \hat{I}(t) \rangle  \\
\langle \delta \hat{I}(t') \delta \hat{I}(t) \rangle & = & \delta(t-t') \langle \hat{I}(t) \rangle -\langle \hat{I}(t') \rangle \langle \hat{I}(t)\rangle
\end{eqnarray}
Considering an exponential dependence of the average current, $\hat{I} = \frac{e}{\tau_e} e^{-t/\tau_e}$, the noise spectrum can be explicitly computed \cite{Mahe2010}:
\begin{eqnarray}
S(\omega) & = &   2 e^2 f \Big[1 - \frac{1}{1 + \omega^2 \tau_e^2}  \Big] = 2e^2 f \frac{\omega^2 \tau_e^2}{1 + \omega^2 \tau_e^2} \label{EqSom}
\end{eqnarray}
This result has been obtained using a semiclassical stochastic model \cite{Albert2010} of single electron emission from a dot containing a single electron. This model also does not take into account correlations with the Fermi sea. Note also that our single electron emitter generates one electron followed by one hole in a period $T_0$, that is one charge in time $T_0/2$.  The factor $2 e^2 f$ in Eq.(\ref{EqSom}) needs to be replaced by $4e^2 f$ in our case. A typical trace for the noise spectrum is plotted on Fig.\ref{fig3_2}. The noise vanishes at low frequency and grows on a scale given by the average escape time, $\omega \sim 1/\tau_e$. To reveal single particle emission, current correlations need to be measured on a time scale shorter than the average escape time, that is, through high frequency noise measurements (typically at GHz frequencies) \cite{Zakka-Bajjani2007}. As exactly a single particle is emitted at each cycle of the source, the fluctuations cannot be attributed to fluctuations in the emitted charge but rather to fluctuations in the emission time. Due to the tunneling emission process, there is a random jitter between the emission trigger and the emission time. Following Eq.(\ref{EqSom}), the noise goes to a white noise limit at high frequency $\omega \tau_e \gg 1$ where correlations are dominated by the first term proportional to $\delta(t-t')$ \cite{Parmentier2010,Moskalets2013}. However at these high frequencies, correlations with the Fermi sea, Eq.(\ref{EqSFS}) cannot be neglected and are responsible for a high frequency cutoff of the noise when $\omega \geq \Delta/(2 \hbar)$ (correlations with the Fermi sea are plotted on blue dashed line on Fig.\ref{fig3_2}. Indeed this cutoff can be interpreted as the impossibility for a particle of energy $\Delta/2$ above the Fermi sea to emit a photon of energy greater than $\hbar \omega = \Delta/2$ due to Pauli blocking by the Fermi sea. A good choice of the measurement frequency thus lies between these two limits :  $\frac{1}{\tau_e} \approx \omega \ll \Delta / \hbar$ which naturally sets the GHz as the appropriate range.

\subsection{High frequency noise measurements}

\begin{figure}[h!]
\centering\includegraphics[width=0.5\textwidth]{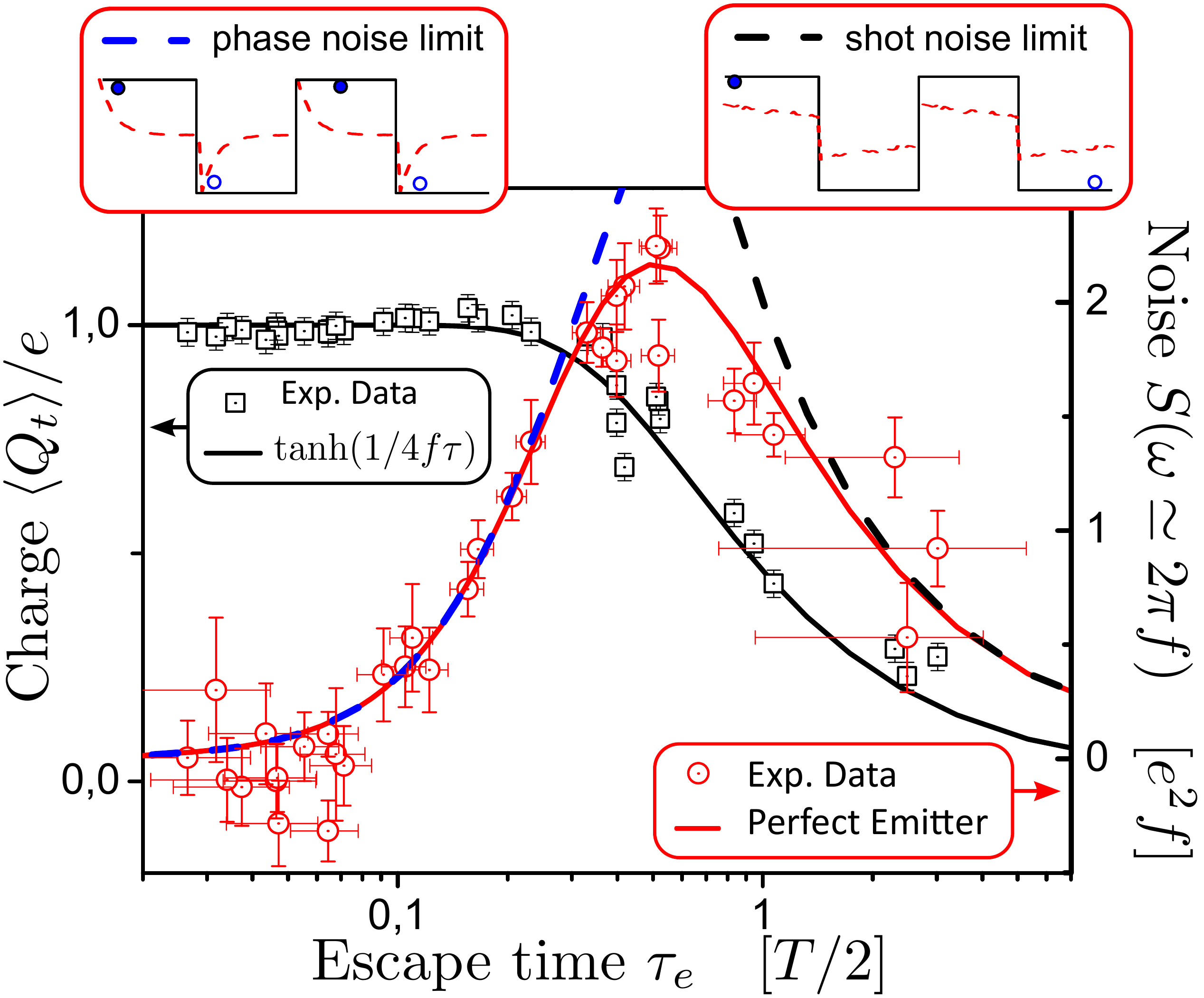}
\caption{\label{fig3_4} Measurements of the high frequency noise $S(\omega=2 \pi f)$, red dots, as a function of the emission time $\tau_e$. The red trace corresponds to the expected dependence, Eq.(\ref{EqS}). Dashed lines correspond to the asymptotic limits of perfect emitter (blue dashed line) and shot noise (black dashed line). The black points correspond to the measurements of the emission probability.
}
\end{figure}

 In the noise measurement, the output ohmic contact on Figure \ref{fig2_1} is used both for the determination of the average current and the high frequency noise (for further experimental details, see ref. \cite{Parmentier2011}). The typical order of magnitude for the noise is given by $e^2f \approx 4. 10^{-29} A^2.Hz^{-1}$ for a drive frequency $f\approx 1.5 GHz$. We implemented a high frequency noise measurement with a 600 MHz bandwidth centered on the drive frequency and a noise sensitivity of a few $10^{-30} A^2.Hz^{-1}$ in a few hours measurement time.  The noise was calibrated by measuring the equilibrium noise of a $50$ Ohms resistor as a function of the temperature. In such noise measurements, it is very hard to change the measurement frequency as it would be required in order to check Eq.(\ref{EqSom}). However, the dependence in the measurement frequency goes like $\omega \tau_e$ which allows to work at fixed frequency, chosen as $\omega=2 \pi f$ (where $f$ is the frequency of the excitation drive) but variable average escape time to check the frequency dependence. Measurements of the noise \cite{Mahe2010, Parmentier2012} as a function of the escape time are plotted on Fig. \ref{fig3_4}. For short escape times, the noise exactly follows the expected dependence (blue trace). However, when the escape time becomes comparable with the half period, the noise deviates from the limit of the perfect emitter. This can be understood, as in this limit of long escape times, electrons do not have enough time to escape the dot and the probability of single charge emission deviates from $1$ (black dots on Fig. \ref{fig3_4}). For an average current following an exponential dependence, the probability $P$ can be computed as a function of the average escape time, $P= \tanh{T_0/4\tau_e}$. As can be seen on Fig. \ref{fig3_4}, the experimental points fall precisely on this $\tanh{T_0/4\tau_e}$ dependence (black trace). This finite probability of charge emission has been taken into account in the heuristic semiclassical model \cite{Albert2010, Parmentier2012} of single charge emission mentioned above, the perfect emitter formula is then modified in the following way:
\begin{eqnarray}
S(\omega) & = &   4 e^2 f \tanh{\frac{T_0}{4\tau_e}} \frac{\omega^2 \tau_e^2}{1 + \omega^2 \tau_e^2} \label{EqS}
\end{eqnarray}
This dependence of the noise for an arbitrary value of the dot transmission can also be confirmed by numerical simulations within the Floquet scattering formalism \cite{Parmentier2012} described above or by real time calculations of single charge emission in a tight-binding model \cite{Jonckheere2012a}. Our data points agree remarkably well with this dependence (red trace) which defines two limits. For short times, the noise follows the perfect emitter limit, there are no fluctuations in the emitted charge and the noise is governed by the random jitter in the emission time. In the long time limit, the fluctuations are governed by the fluctuations in the number of emitted charges. Taking $\omega \tau_e \gg 1$ in Eq.(\ref{EqS}), the noise becomes independent of frequency and proportional to the average current, $S(\omega) \approx 2 e |\langle \hat{I}(t) \rangle |$ for $\tau_e \gg T_0/2$. In this limit single charge emission becomes a random poissonian process. Figure \ref{fig3_4} shows the proper conditions to operate the source as a good single particle emitter, for $\tau_e\leq 0.3 T_0/2$, the source follows the perfect emitter limit.

To conclude this section, average current measurement of a triggered electron emitter show that the source emits on average a quantized number of particles. The measurement of second order correlations can then be used to demonstrate that a single particle is emitted at each emission cycle. This single electron emitter will then be used to characterize and manipulate single electron states in optics-like setups. In particular, the Hanbury-Brown and Twiss geometry, where the electron beams are partitioned by a beam-splitter will be thoroughly studied.

\section{Hanbury-Brown \& Twiss interferometry}
\label{HBT}

\begin{figure}
  \includegraphics[width=\columnwidth]{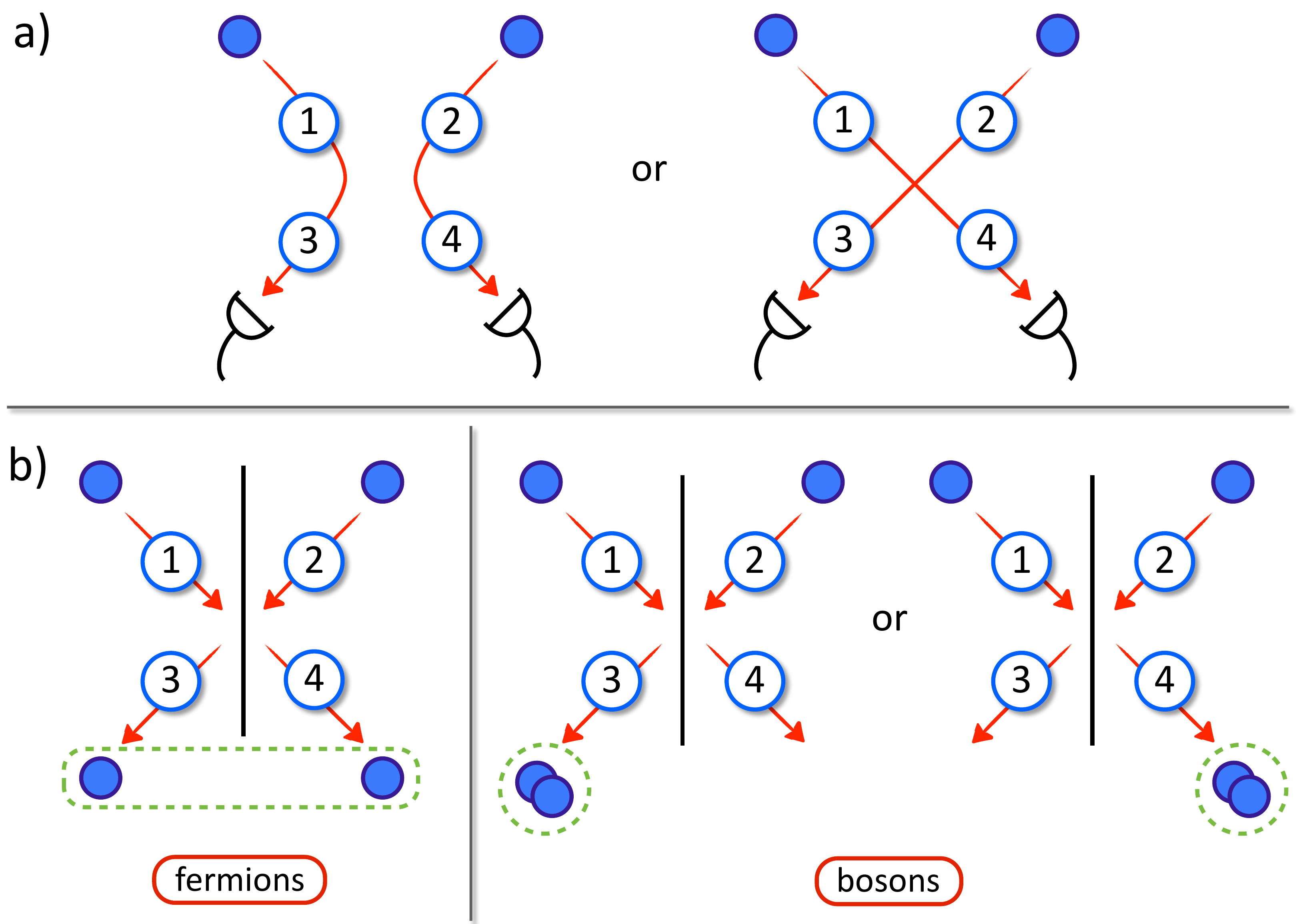}
  \caption{\label{Fig:Antibunching}
a) Direct and exchange paths, that interfere when placing two sources in inputs 1 and 2 and recording correlations between two detectors at  outputs 3 and 4. b) Possible outcomes of two-particle interference experiments when two indistinguishable particles are placed in the inputs of a beamsplitter.}
\end{figure}

When studying the correlations between two sources using two detectors, the Hanbury-Brown \& Twiss effect arises from two-particle interferences between direct and exchange paths, pictured on Fig.\ref{Fig:Antibunching} a). As discovered in 1956 when observing distant stars \cite{Hanbury-Brown1956}, intensity correlations offer a powerful way to study the emission statistics of sources. In particular, two particle interferences lead to different possible outcomes depending on the fermionic or bosonic character of the two indistinguishable particles that would impinge on a beamsplitter (Fig. \ref{Fig:Antibunching} b)). On one hand, indistinguishable electrons (fermions) antibunch: the only possible outcome is to measure one electron in each output arm. On the other hand, indistinguishable photons (bosons) bunch: two photons are then measured in one of the outputs. Thus, when such particles collide and bunch/antibunch on the beam-splitter, the fluctuations and correlations of output currents encode information on the single particle content of the incoming beams. First observed with light sources \cite{Hanbury-Brown1957}, the HBT effect has since then been observed for electrons propagating in a two dimensional electron gas \cite{Oliver1999,Henny1999,Henny1999a}.

\begin{figure*}
\begin{center}
  \includegraphics[width=0.9\textwidth]{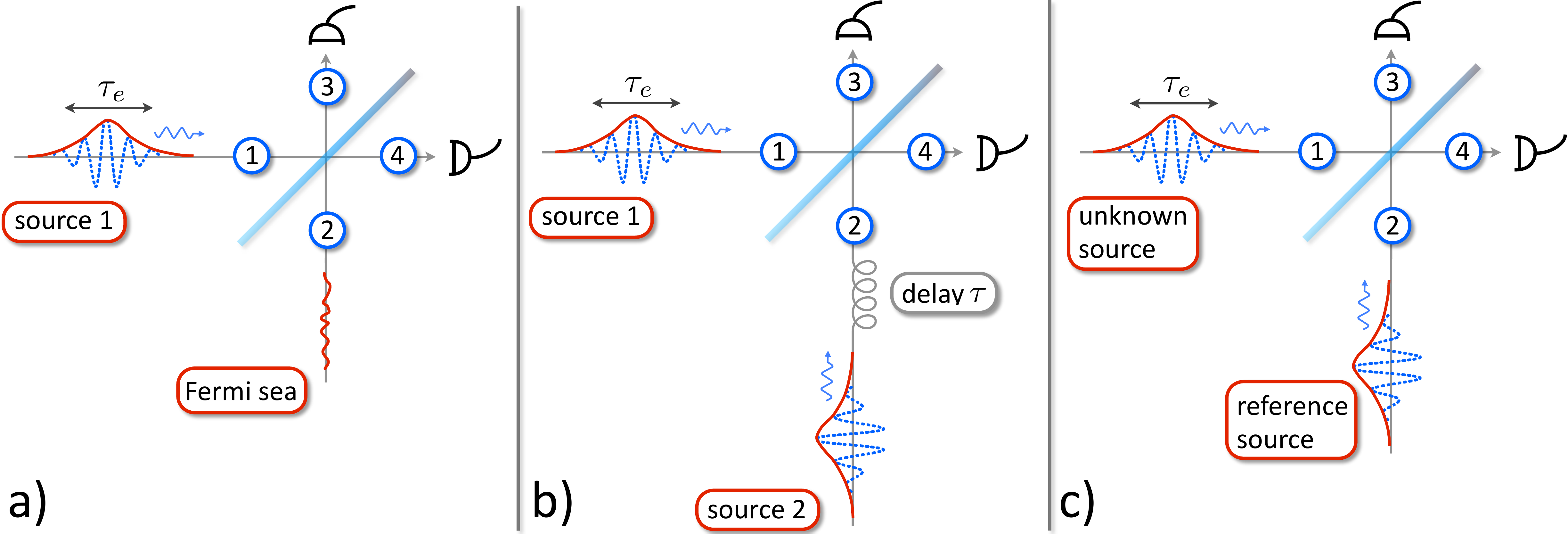}
  \end{center}
  \caption{\label{Fig:HBTExperiments}
    The Hanbury-Brown \& Twiss geometry consists in the measurements of intensity auto- or cross- correlations at the outputs of a beam-splitter (outputs 3 and 4). Depending on the sources (inputs 1 and 2), different properties can be inferred. The source under study (source 1) is plugged in input 1. Three cases corresponding to three different sources connected to the second input are considered in this article: a) source 2 is a Fermi sea ("vacuum") and a single source is partitioned on the splitter, b) source 2 is identical to the one in 1 and the setup is analogous to the optical Hong-Ou-Mandel experiment, c) source 2 is a reference source used in a tomography protocol of source 1.}
\end{figure*}

A convenient way to implement the interference between the two exchanged paths on two detectors is to use the geometry described on Fig.\ref{Fig:HBTExperiments}. The two sources are placed at the two inputs of a beam-splitter and the two detectors at the two outputs. A coincidence detection event on the detectors has then two exchanged contributions. Particles emitted by source 1 and 2 can be reflected to 3 and 4 or transmitted to 4 and 3. These two paths lead to two-particle interferences in the coincidence counts of the two detectors. Using electron sources, a quantum point contact can be used as a tunable electronic beam-splitter with energy-independent reflexion and transmission coefficients $R$ and $T$ ($R+T=1$) relating incoming to outgoing modes. As single particle detection is not available yet for electrons (at least for subnanosecond time scales), coincidence counts are replaced in electronics by current correlations. The output current operators $\hat{I}_\alpha(t),\ (\alpha\in \{3,4\})$ and the output current correlations $S_{\alpha\beta}(t',t)=\langle\delta\hat{I}_\alpha(t')\delta\hat{I}_\beta(t)\rangle,\ (\alpha,\beta\in\{3,4\})$ can be expressed in terms of input currents and correlations :
\begin{eqnarray}
S_{33}(t',t)&=&R^2 S_{11}(t',t)+T^2 S_{22}(t',t)+RT Q(t,t') \quad  \label{S33} \\
S_{44}(t',t)&=&T^2 S_{11}(t',t)+R^2 S_{22}(t',t)+RT Q(t,t') \quad \label{S44} \\
S_{34}(t',t)&=&RT \big(S_{11}(t',t)+S_{22}(t',t)- Q(t,t')) \quad \label{S34}
\end{eqnarray}
where $S_{11}(t',t)$ and $S_{22}(t',t)$ are the current fluctuations in inputs 1 and 2 and $Q(t,t')$ denotes the quantum Hanbury-Brown \& Twiss contribution to outcoming current correlations. It encodes the aforementioned two-particle interferences and involves the coherence functions of incoming electrons and holes :
\begin{eqnarray}
Q(t,t')&=& e^2 v^2 \;  \mathcal{G}_1^{(1,e)}(t,t')\mathcal{G}_2^{(1,h)}(t,t') \nonumber \\
 & + & e^2 v^2 \;\mathcal{G}_1^{(1,h)}(t,t')\mathcal{G}_2^{(1,e)}(t,t')
\end{eqnarray}
This quantum two-particle interference can be unveiled through the measurement of zero-frequency correlations. Namely, standard low-frequency noise measurement setup gives access to the averaged quantities $S_{\alpha\beta}(\omega=0)=2\int d\tau\ \overline{S_{\alpha\beta}(t + \tau,t)^t}$.
Thus it is possible to access the averaged HBT contribution
\begin{eqnarray}
\overline Q&=&2 e^2 v^2\int d\tau\ \big[ \; \overline{\mathcal{G}_1^{(1,e)}(t,t+\tau)\mathcal{G}_2^{(1,h)}(t,t+\tau)^t}  \nonumber \\
&  &\quad + \; \overline{\mathcal{G}_1^{(1,h)}(t,t+\tau)\mathcal{G}_1^{(1,e)}(t,t+\tau)^t} \big] \label{Eq:Qbar}
\end{eqnarray}
which is nothing but the overlap between the single electron and hole coherences of channels 1 and 2, and plays a key role in the various experiments one can perform in the Hanbury-Brown \& Twiss geometry.
In the following, we will study the three situations described on Fig.\ref{Fig:HBTExperiments}. In the first one, a single source is used and partitioned on the splitter while the second input is kept 'empty'. Contrary to the true vacuum obtained in the optical experiment, in electronics, this second input is always connected to a Fermi sea which is a source at equilibrium. This leads to important differences in the electronic version of this experiment. In the second experiment, each input is connected to a triggered single electron emitter. Two single electrons collide synchronously on the splitter realizing the electronic analog of the Hong-Ou-Mandel experiment in optics \cite{Hong1987, Santori2002,Beugnon2006, Lang2013}. Finally, using a reference state in one input, an unknown input state can be reconstructed and imaged by measuring its overlap with the known reference state. The principle of such a single electron state tomography will be described in the last section.

\subsection{Single source partitioning}

Let us first consider the electronic analog of the seminal experiment performed by Hanbury-Brown \& Twiss to characterize optical sources \cite{Hanbury-Brown1957}, in which a light source is placed in input 1 whereas the second arm is empty and described by the vacuum. In the electronic analog, the single electron source described previously is used, while the empty arm now consists of a Fermi sea at equilibrium, with fixed temperature and chemical potential.
The purpose of this experiment is not here to obtain the charge statistics of the source, that is accessed via high-frequency autocorrelations described in the previous section. It in fact reveals the number of elementary excitations (electron/hole pairs) produced by the electron source, which has no optical counterpart and stems from the fact that particles with opposite charges contribute with opposite signs to the current. The total number of elementary excitations emitted from the source is hard to access through a direct measurement of the current or its correlations (that is without partitioning). Indeed, the emission of one additional spurious electron/hole pair in one driving period, as represented on Fig.\ref{fig3_1} (sixth period of the drive on the figure) is a neutral process and cannot be revealed in the current if the time resolution of the current measurement is longer that the temporal separation between the electron and the hole. This temporal resolution is estimated to be a few tens of picoseconds in the high frequency noise measurement presented previously. Spurious electron/hole pairs emitted by the source on a shorter time scale thus cannot be detected. However, the random and independent partitioning of electrons and holes on the splitter can be used to deduce their number from the low frequency current fluctuations of the output currents. Using Eqs.(\ref{S44}-\ref{Eq:Qbar}), the excess output current correlations and their low frequency spectrum are given by :
\begin{eqnarray}
\Delta S_{33}(t',t) &=& \Delta S_{44}(t',t) = - \Delta S_{34}(t',t) = R T \Delta Q(t,t') \nonumber \\
\\
S_{33}(\omega=0) &=&  RT \Delta \overline{Q} \\
  & =& 2 RT e^2v^2 \int d\tau \Big[ \overline{\Delta \mathcal{G}_1^{(1,e)}(t,t+\tau)}^{t} \mathcal{G}_F^{(1,h)}(\tau) \nonumber \\
& & \quad \quad \quad \quad + \; \overline{\Delta \mathcal{G}^{(1,h)}_1(t,t+\tau)}^{t} \mathcal{G}_F^{(1,e)}(\tau) \Big] \label{Eq:DeltaQbar}
\end{eqnarray}
Where $\Delta Q(t,t')$ is the excess HBT contribution with respect to equilibrium. As can be seen in Eq.(\ref{Eq:DeltaQbar}) and contrary to optics, the single source partitioning experiment involves two sources, the triggered emitter and the Fermi sea at finite temperature, through the overlap between their first order coherence $\Delta \mathcal{G}^{(1)}_1$ and $\mathcal{G}_F^{(1)}$. This overlap is more easily expressed in Fourier space:
\begin{eqnarray}
\Delta\overline Q&=& 2 \frac{e^2}{T_{meas}} \int_{0}^{+\infty} d\epsilon\ \big[ \delta n_e(\epsilon) + \delta n_h(\epsilon) \big] \big(1-2 f(\epsilon)\big) \nonumber \\ \label{Eq:DeltaQHBT}
\end{eqnarray}
Where $\delta n_e(\epsilon)$ is the excess number of electrons (at energy $\epsilon \geq 0$ above the Fermi energy) emitted per unit energy in the long measurement time $T_{meas}$. Similarly, $\delta n_h(\epsilon)$ is the energy density of the number of holes emitted at energy $\epsilon \geq 0$ (corresponding to a missing electron at energy $-\epsilon$ below the Fermi energy) in the measurement time $T_{meas}$. For a periodic emitter of frequency $f$, it is more convenient to use the energy density of the number of excitations emitted in one period. To avoid defining too many notations, in the rest of the manuscript, $\delta n_e(\epsilon)$ (resp. $\delta n_h(\epsilon)$) will refer to the energy density of electrons (resp. holes) emitted in one period. Defining $\delta N_{HBT}$ as the number of electron/hole pairs counted per period by the partition noise measurement, Eq.(\ref{Eq:DeltaQHBT}) then becomes:
\begin{eqnarray}
\Delta\overline Q&=& 4 e^2 f \delta N_{HBT} \\
\delta N_{HBT} & = &  \int_{0}^{+\infty} d\epsilon\ \frac{\delta n_e(\epsilon) + \delta n_h(\epsilon)}{2} \big(1-2 f(\epsilon)\big)
\end{eqnarray}

Considering first the limit of zero temperature, $\delta N_{HBT}=\frac{\langle \delta N_e \rangle + \langle \delta N_h \rangle }{2}$ equals the average number of electrons/holes emitted in one period. This result can be understood by a simple classical reasoning: electrons and holes are independently partitioned on the beam-splitter following a binomial law. As a consequence, the low-frequency output noise is proportional to the number of elementary excitations arriving on the splitter. Consequently, measuring the HBT contribution directly gives access to the total number of excitations generated per emission cycle. However, large deviations to this classical result can be observed due to finite temperature. Indeed, input arms are populated with thermal electron/hole excitations that can interfere with the ones generated by the source, thus affecting their partitioning. As seen in Eq. \ref{Eq:DeltaQHBT}, $\delta N_{HBT}$ is corrected by $-\int d\epsilon \big(\delta n_e(\epsilon)+\delta n_h(\epsilon)\big)f(\epsilon)$, corresponding to the energy overlap of thermal excitations and the particles triggered by the source. The minus sign reflects the fermionic nature of particles colliding on the QPC. For vanishing temperatures, classical partitioning is recovered. For non-vanishing temperature, a fraction of the triggered excitations reaching the beamsplitter find thermal ones at the same energy. In virtue of Fermi-Dirac statistics, these indistinguishable excitations antibunch (see Fig.\ref{Fig:Antibunching}): the only possible outcome consists of one excitation in each output, so that no fluctuations are expected in that case, thus reducing the amplitude of the HBT correlations.

\begin{figure}
  \includegraphics[width=\columnwidth]{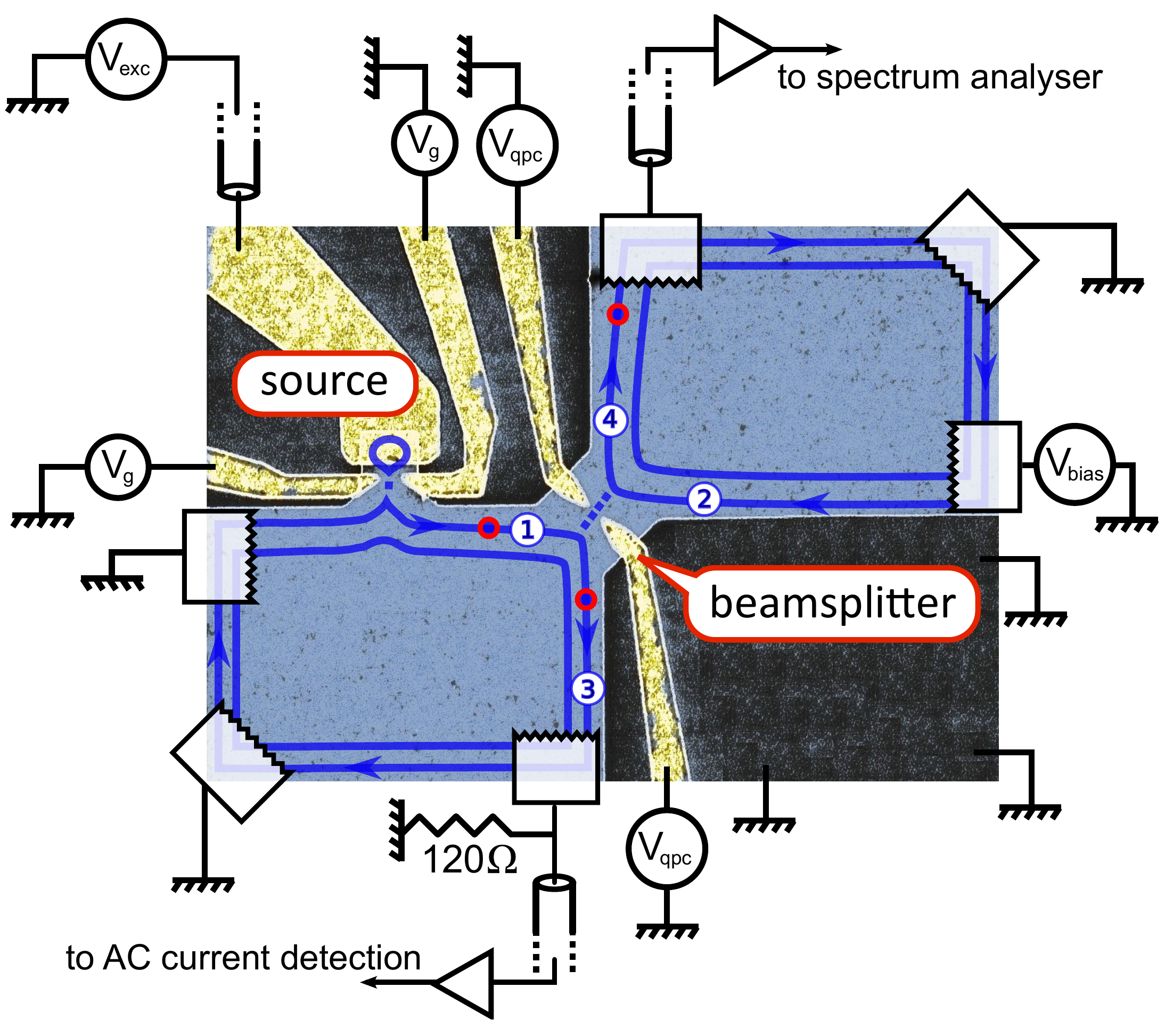}
  \caption{\label{Fig:HBTSample}
    Modified SEM picture of the sample used in the Hanbury-Brown \& Twiss experiment. A perpendicular magnetic field $B=3.2$ T is applied in order to work at filling factor $\nu=2$. The two edge channels are represented by blue lines. The emitter is placed on input $1$, 2.5 microns before the electronic splitter whose gate voltage $V_{qpc}$ is set to fully reflect the inner edge while the outer edge can be partially transmitted with tuneable transmission $T$.  The emitter is tunnel coupled to the outer edge channel with a transmission $D$ tuned by the gate voltage $V_g$. Electron emission is triggered by the excitation drive $V_{exc}(t)$. Average measurements of the AC current generated by the source are performed on output $3$, whereas output $4$ is dedicated to the low frequency noise measurements $S_{4,4}$.}
\end{figure}

An experimental realization \cite{Bocquillon2012} confirms these findings. The single electron emitter described in the previous section \ref{CapaMeso} is placed on input 1 of a quantum point contact (at a distance of approximately 3 microns), see Fig.\ref{Fig:HBTSample}. Low frequency current correlations $S_{44}$ are measured on output $4$ while output $3$ is used to to characterize the source through high frequency measurements of the average ac current generated by the source. The emitter is driven at a frequency of 1.7 GHz with different excitation drives (sine or square waves) so as to generate different wavepackets. For  transmissions $0.2<D<0.7$, the average emitted charge $\langle Q_t\rangle$ deduced from measurements of the average ac current equals the elementary charge $e$ with an accuracy of 10 \%. For $D\simeq1$, $\langle Q_t\rangle$ exceeds $e$ as quantization effects in the dot vanish, and $\langle Q_t\rangle\to0$ for $D\to 0$.

\begin{figure}
  \includegraphics[width=\columnwidth]{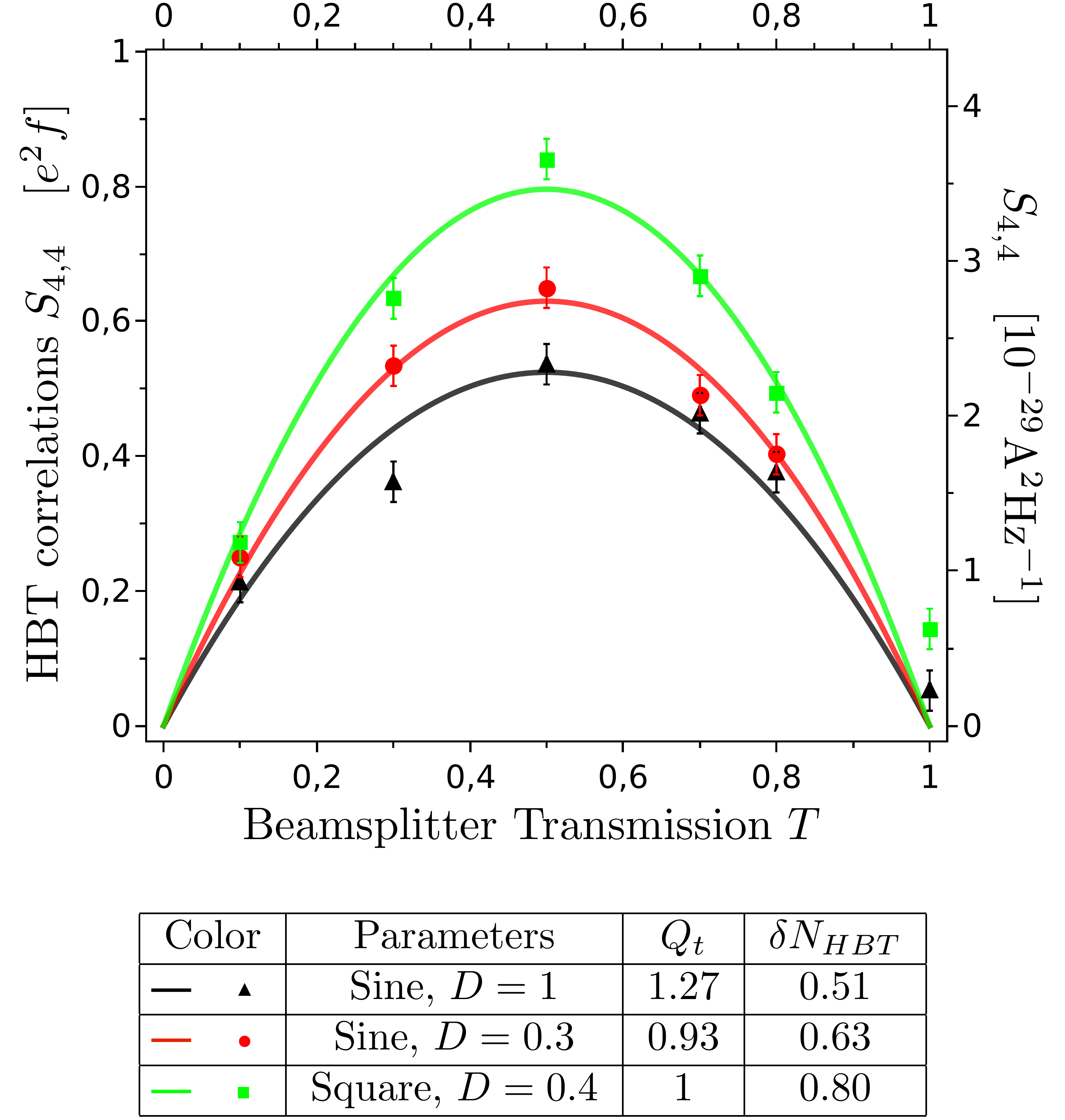}
  \caption{\label{Fig:HBTTdependence}
    Low frequency HBT correlation $S_{44}$ as a function of the transmission of the beamsplitter $T$, in units $e^2f$ (left axis) and in $\rm A^2Hz^{-1}$. Three different rf drives are presented : sine drive at transmission $D=1$ (black triangles), sine drive at transmission $D=0.3$ (red dots), square drive at transmission $D=0.4$ (green squares). The plain lines represent fits with the expected $T(1-T)$ dependence. Different amplitudes of noise are obtained, reflecting the fact that antibunching with thermal excitations strongly depends on the energy distribution of the generated wavepackets.}
\end{figure}

Fig.\ref{Fig:HBTTdependence} presents the HBT low frequency correlations as a function of the beam-splitter transmission $T$. For all three curves, the $T(1-T)$ dependence is observed, but the noise magnitude notably differ. In particular, $\delta N_{HBT}<\langle Q\rangle$, invalidating the classical partitioning of a single electron/hole pair. This discrepancy is attributed to the non-zero overlap between triggered excitations and thermal ones, whose exact value strongly depends on the driving parameters. An intuitive picture can be proposed. The highest value of $\delta N_{HBT}$ is observed with a square drive. In this case, a single energy level in the dot is rapidly raised from below to above the Fermi level of the reservoir, and the quasiparticle is emitted at an energy $\epsilon_e \simeq\frac{\Delta}{2}>k_B T_{el}$ well separated from thermal excitations. Therefore, we expect the outcoming noise to be maximum. For a sine wave, the rise of the energy level in the dot is slower and the electron is emitted at lower energies and thus more prone to antibunch with thermal excitations. This tends to reduce $\delta N_{HBT}$. As the transmission $D$ is lowered, the escape time $\tau_e$ increases and electron emission occurs at later times, corresponding to higher levels of the sine drive. The quasiparticle is then emitted at higher energies and are less sensitive to thermal excitations. $\delta N_{HBT}$ is then increased, as seen by comparing the black and red traces of Fig.\ref{Fig:HBTTdependence}. This intuitive picture can be confronted to numerical calculations within the Floquet scattering theory \cite{Moskalets2008, Parmentier2012} which can be used to calculate $\delta n_e(\epsilon)$ and $\delta n_h(\epsilon)$ for any type of excitation drive (sine or square) and any value of the dot parameters. The resulting curves for the energy distributions can be found on ref \cite{Bocquillon2012}, they confirm the intuitive picture discussed above.

\begin{figure*}
  \includegraphics[width=0.7\textwidth]{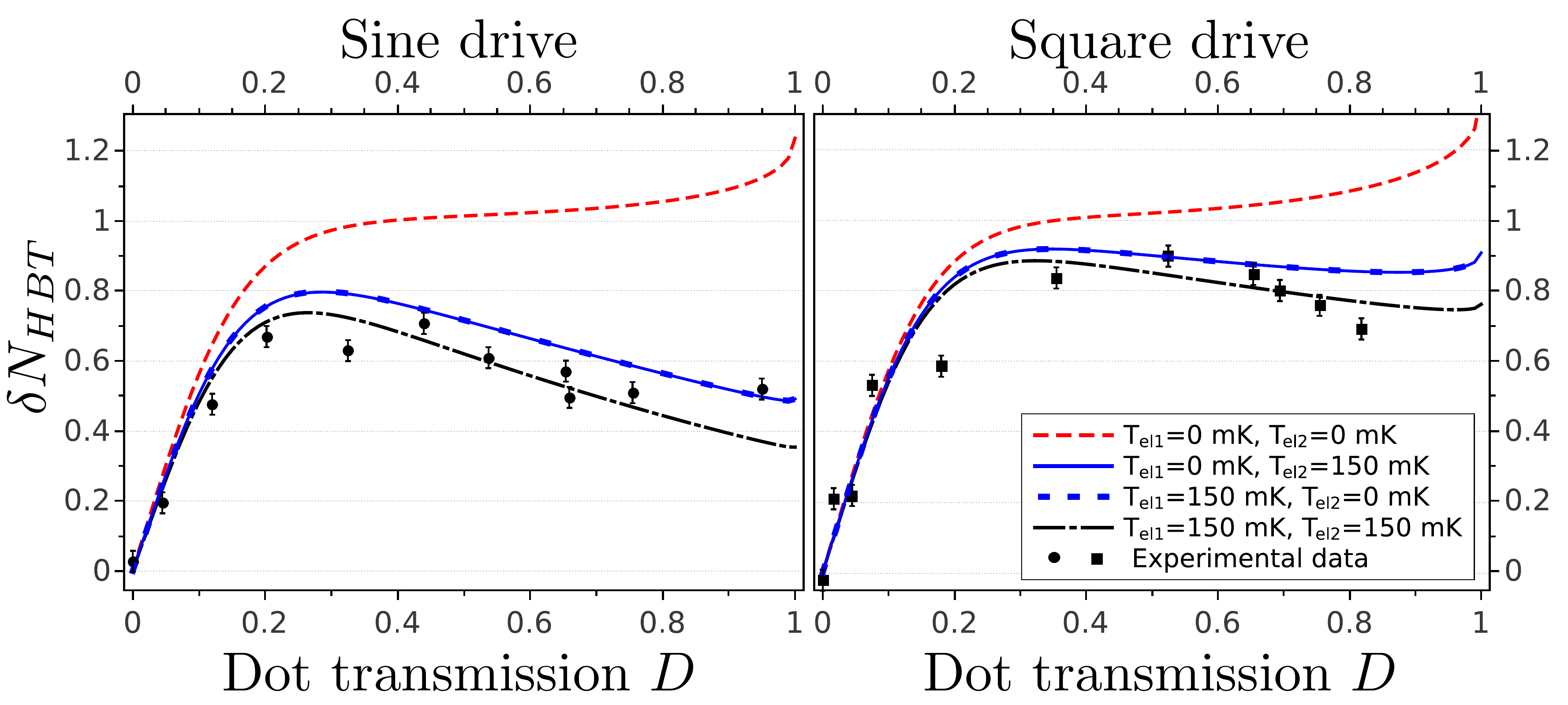}
  \caption{\label{Fig:HBTDdependence}
    HBT contribution $\delta N_{HBT}$ as a function of the dot transmission $D$ for sine drive (left panel) and a square drive (right panel). Experimental points are represented by dots (sine) and squares (square drive) and compared with numerical simulations based on Floquet scattering theory: $T_{el,1}=T_{el,2}=0$ (red dashes), $T_{el,1}=T_{el,2}=150$ mK (black dashes), and $T_{el,1}=150$ mK, $T_{el,2}=0$ and $T_{el,1}=0$, $T_{el,2}=150$ mK (blue plain and dashed lines).}
\end{figure*}

These differences in energy distributions can be revealed by the Hanbury-Brown \& Twiss interferometry, as shown on Fig.\ref{Fig:HBTDdependence} that presents measurements of $\delta N_{HBT}$ as a function of the dot transmission $D$ for two different drives, sine or square. Floquet calculations for square and sine drives at $T_{el}=0$ are presented in red dashed line: they are almost identical and reach $\delta N_{HBT}\simeq 1$ for $D\in[0.2,0.7]$, as expected for an ideal source that does not emit additional electron-hole pairs. For $D<0.2$, the shot noise regime is recovered whereas quantization effects in the dot are progressively lost for $D>0.7$. The effect of temperature in arm 2 ($T_{el,2}=150$ mK, $T_{el,2}=0$) is shown in blue line. As already discussed, the presence of thermal excitations reduces $\delta N_{HBT}$. This effect decreases when lowering the transmission, and is more pronounced for sine wave than for square drive. Remarkably, the effect of temperature in arm 1 (blue dashes) is identical to the one in arm 2. When a temperature of 150 mK (extracted from noise thermometry) is introduced in both arms, a good agreement is found with the experimental data (black dashes). This confirms the tendency to produce low energy excitations when using a sine drive, and energy-resolved excitations using a square drive. Note that the Floquet calculations do not take into account the energy relaxation \cite{Degiovanni2009} along the 3 microns propagation towards the splitter that will be discussed in the last section of this article. It only provides the energy distribution at the output of the source, 3 microns away from the splitter where the collision with thermal excitations occur. The good agreement with Floquet calculation implies that energy relaxation has a small effect on the total number of excitations and would require a direct measurement of the energy distribution (and not of its integral on all energies) to be characterized.

\subsection{Hong-Ou-Mandel experiment}
\label{HBThom}
\begin{figure}
  \includegraphics[width=\columnwidth]{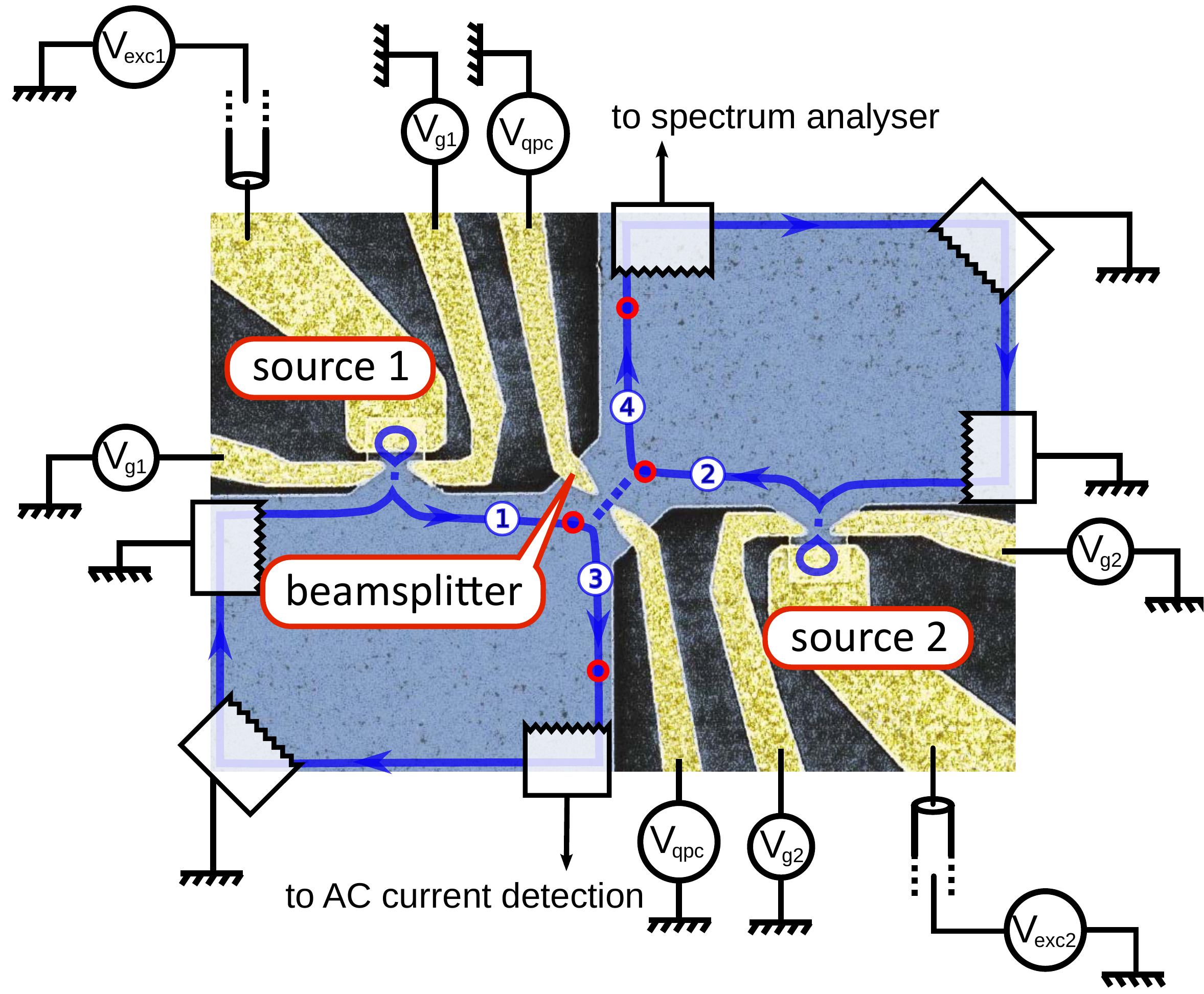}
  \caption{\label{Fig:HOMSample}
   Modified SEM picture of the sample used in the Hong-Ou-Mandel experiment. The electron gas is represented in blue. Two single-electron emitters are located at inputs $1$ and $2$ of a quantum point contact used as a single electron beamsplitter. Transparencies $D_1$ and $D_2$ and static potentials of dots $1$ and $2$ are tuned by gate voltages $V_{g,1}$ and $V_{g,2}$. Electron/hole emissions are triggered by excitation drives $V_{exc,1}$ and $V_{exc,2}$. The transparency of the beamsplitter partitioning the inner edge channel (blue line) is tuned by gate voltage $V_{qpc}$ and set at $T=1/2$. The average ac current generated by sources $1$ and $2$ are measured on output 3 while the low frequency output noise $S_{44}$ is measured on output $4$.}
\end{figure}

The previously discussed antibunching effect bears strong analogies with the photon coalescence observed in the Hong-Ou-Mandel experiment \cite{Hong1987}. While quasiparticles are generated on-demand in the first input, thermal excitations are however randomly emitted in the second input. To recreate the electronic analog of the seminal Hong-Ou-Mandel experiment \cite{Feve2008,Olkhovskaya2008,Jonckheere2012}, two identical but independent single electron sources can be placed in the two input arms of the beamsplitter, as pictured in Fig. \ref{Fig:HOMSample}.

As in the seminal HOM experiment, the antibunching of the on-demand quasiparticles provides a direct measurement of the overlap of the two mono-electronic wavefunctions, i.e. their degree of indistinguishability. Indeed, for two sources generating periodically (period $1/f$) a single electron described by the wavefunctions $\phi_1^{e}(x)$ and $\phi_2^{e}(x)$ above the Fermi sea (well separated from thermal excitations), as seen in section \ref{StationarySES}, the coherence function for source i reads $\Delta \mathcal{G}_i^{(1,e)}(t,t') = \phi_i^{e}(-vt) \phi_i^{e,*}(-vt')$ such that we have:
\begin{eqnarray}
\Delta\overline Q&=& 4e^2f\big(1-|\langle\phi_1^{e}|\phi_2^{e}\rangle|^2\big)
\end{eqnarray}
For perfectly distinguishable electrons, $\langle\phi_1|\phi_2\rangle=0$ and the classical random partitioning of two electrons is recovered. However, for perfectly indistinguishable electrons, $\langle\phi_1|\phi_2\rangle=1$ and the random partitioning is fully suppressed. The overlap between the two particles can be modulated by varying the delay $\tau$ between the excitations drives. Dividing $\Delta\overline Q$ by the total partition noise of both sources ($2 e^2f$ for each source neglecting temperature effects) one then gets the normalized HOM correlations $\Delta \overline q$ as:
\begin{eqnarray}
\Delta\overline q&=&1-\big|\int dt\  \phi_1^{e,*}(t)\phi_2^{e}(t+\tau)\big|^2\label{eq:HOM}
\end{eqnarray}
When working at finite temperature, the partition noise in the HOM and HBT configurations is reduced from their overlap with thermal excitations (see previous section). However, if the generated quantum states in sources 1 and 2 remain indistinguishable, the antibunching effect remains total and numerical simulations using the Floquet scattering formalism show that $\Delta \overline q$ is only marginally modified.

This experiment \cite{Bocquillon2013a} was realized using similar sources (level spacings $\Delta_1\simeq\Delta_2\simeq 1.4\pm0.1$ K), driven at frequency $f=2.1$ GHz with square waves. A delay $\tau$ between both drives can be tuned with an accuracy of $7$ ps. For $D_1\simeq D_2\simeq 0.4$, both sources are expected to produce energy-resolved excitations relatively well-separated from the Fermi sea and with charge $\langle Q_t \rangle\simeq e$, thus achieving with reasonable accuracy the ideal generation of single-electrons wavepackets.

\begin{figure}
  \includegraphics[width=\columnwidth]{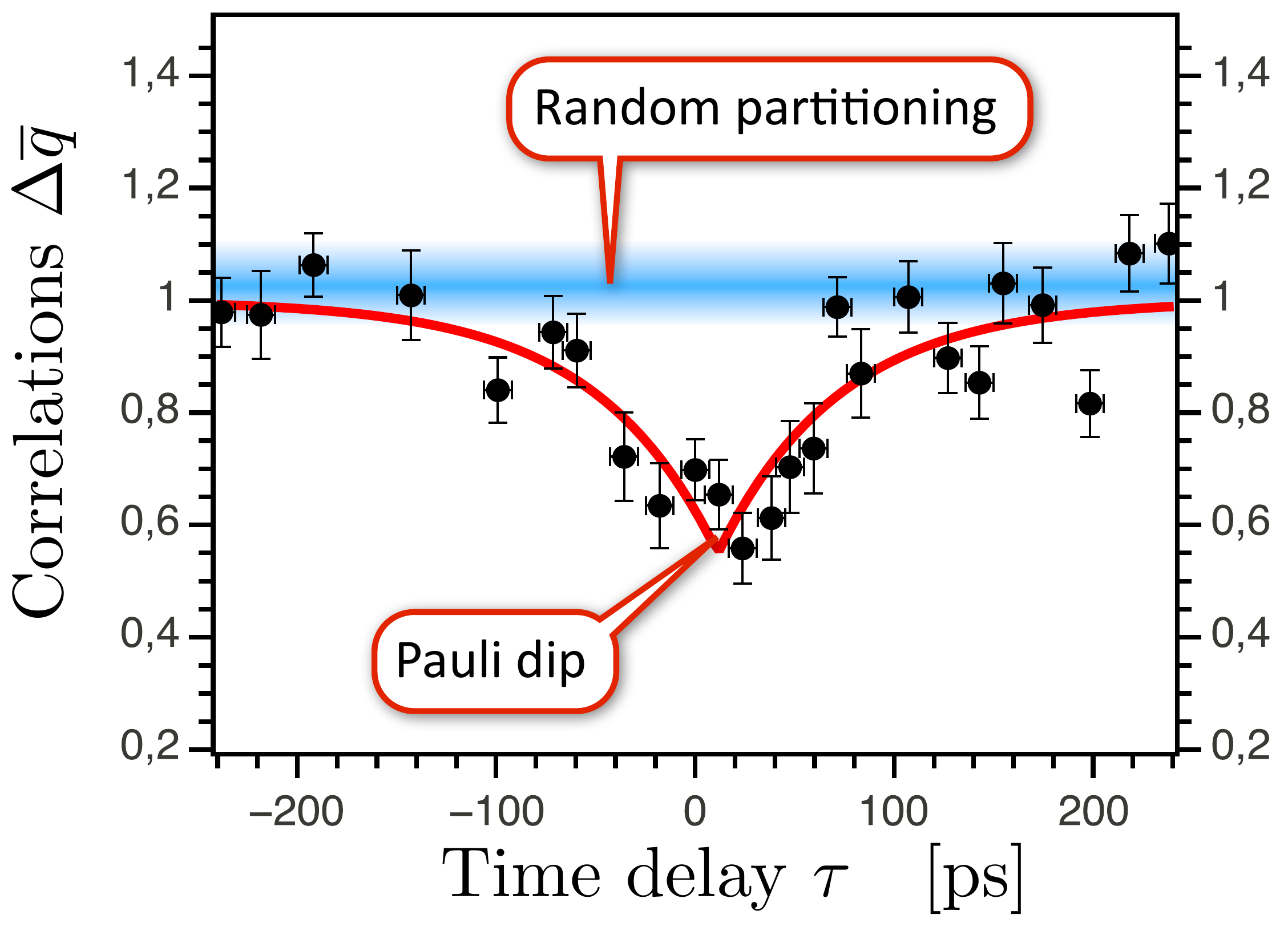}
  \caption{\label{Fig:HOMDip}
    Excess noise $\Delta \overline q$ as a function of time delay $\tau$ and normalized by the value on the plateau observed for long delays. The sum of both partition noises (in the HBT configuration) is depicted by the blue blurry line, while the red trace is obtained with a fit by $\Delta\overline q=1-\eta e^{-|\tau-\tau_0|/\tau_e}$}
\end{figure}

The resulting HOM correlations are presented in Fig.\ref{Fig:HOMDip} as a function of delay $\tau$. A dip in the correlations is clearly observed around $\tau=0$. The measured noise is normalized by its value on the plateaus observed at large delays, and matches as expected the sum of the HBT contributions of each source, that are measured independently by alternatively turning one of the sources off. As seen in section \ref{SES}, for a square wave excitation, single electron emission is described by an exponentially decaying wavepacket, with decay time $\tau_e$ and energy $\epsilon_0$ that depends on the amplitude of the square excitation: $\phi_1^{e}(t)=\phi_2^{e}(t)=\frac{\theta(t)}{\sqrt{\tau_e}}e^{-t/2\tau_e}e^{-i\epsilon_0 t/\hbar}$. $\Delta \overline q$ then takes the following simple form :
\begin{eqnarray}
\Delta\overline q&=&1-e^{-|\tau|/\tau_e}
\end{eqnarray}
Taking into account a loss in the visibility $\eta$ and an error on synchronization $\tau_0$, fitting with $\Delta\overline q=1-\eta e^{-|\tau-\tau_0|/\tau_e}$ then gives $\tau_0\simeq 11$ ps, $\tau_e=62\pm10$ ps and $\eta=0.5$. The extracted value of $\tau_e$ is consistent with independent measurements via the average current. Though effects of the partial indistinguishability of the generated excitations are indubitable, the visibility $\eta$ is far from unity. This may be the result of parameter mismatch between the two sources, resulting in reduced overlap of the wavepackets, but also from decoherence effects due to interaction with the environment. Such effects will be discussed in section \ref{Interactions}.\\

\subsection{Electron-hole correlations in the Hong-Ou-Mandel setup}
A unique property of electron optics compared to photon optics is the ability to manipulate hole excitations in addition to electron excitations.
Performing the HOM experiment with identical single hole excitations in the two input arms of the beamsplitter will produce results similar to
those of electrons (with hole wavefunctions replacing electron wavefunctions in Eq.(\ref{eq:HOM})). But performing the HOM experiment
while injecting a single electron excitation in one input arm of the beam-splitter, and a single hole excitation in the other arm will
produce results which have no counterpart in optics.\cite{Jonckheere2012}

In order to get useful analytical formulas,
we first consider theoretically states where one electron charge has been added (removed) from the Fermi sea
\begin{align}
| \Psi_e \rangle & = \int \! dx \; \phi^e(x) \, \psi^{\dagger}(x) \, |F \rangle  \nonumber\\
| \Psi_h \rangle & = \int \! dx \; \phi^h(x) \, \psi(x) \, |F \rangle
\end{align}
where $|F \rangle$ is the Fermi sea at temperature $T_{el}$, and $\phi^e(x)$, $\phi^h(x)$ the electron and the hole wavefunctions in real space.
 Taking the electron-hole symmetric case for simplicity ($\phi^e(\epsilon_F + \delta \epsilon) = \phi^h(\epsilon_F - \delta \epsilon)$), the normalized HOM correlation $\Delta \bar{q}$ becomes:
\begin{equation}
\Delta \bar{q} = 1 + \left|\frac{ \int_0^\infty \! d\epsilon \,
  \phi^{e}(\epsilon) \phi^{h, *}(\epsilon) e^{-i \epsilon \tau/\hbar} f(\epsilon) (1-f(\epsilon))}
  {\int_0^\infty \! d\epsilon  \, |\phi^{e}(\epsilon)|^2 (1-f(\epsilon))^2} \right|^2 \; .
\label{eq:HOM_eh}
\end{equation}
Comparing this with Eq.~(\ref{eq:HOM}), we notice important changes.
First, the interferences contribute now with a positive sign to
the HOM correlations, that is, the opposite of the electron-electron case.
Electron-hole interferences produce a``HOM peak'' rather than
a dip. Second, the value of this peak depends on the overlap of
the electron and the hole wave packets times the
Fermi product $f(\epsilon) (1-f(\epsilon))$. This peak thus vanishes as $T_{el} \to 0$
since it requires a significant  overlap between electron and hole
wave packets, a situation which only happens in an energy
range $\sim k_B T_{el}$ around $\epsilon_F$, where electronic states are neither
fully occupied nor empty.

Note that the many-body state $|\Psi_e\rangle$ (or $|\Psi_h\rangle$)
 created by the application of the electron creation (or annihilation) operator
 is quite complex when the wavepacket $\phi^e(x)$ (or $\phi^h(x)$) has an important weight
close to the Fermi energy. Indeed, due to the changes imposed on the Fermi sea,
many electron-hole pairs are created, and the state is not simply one electron (or one
hole) plus the unperturbed Fermi sea.
The appearance of a positive HOM peak can be attributed to interferences between these
electron-hole pairs coming from the two branches of the setup.
  It is quite remarkable that eventually, the peak can simply be computed from
the overlap of the electron and hole wavepackets (see Eq.(\ref{eq:HOM_eh})).

\begin{figure}
\centerline{\includegraphics[width=\columnwidth]{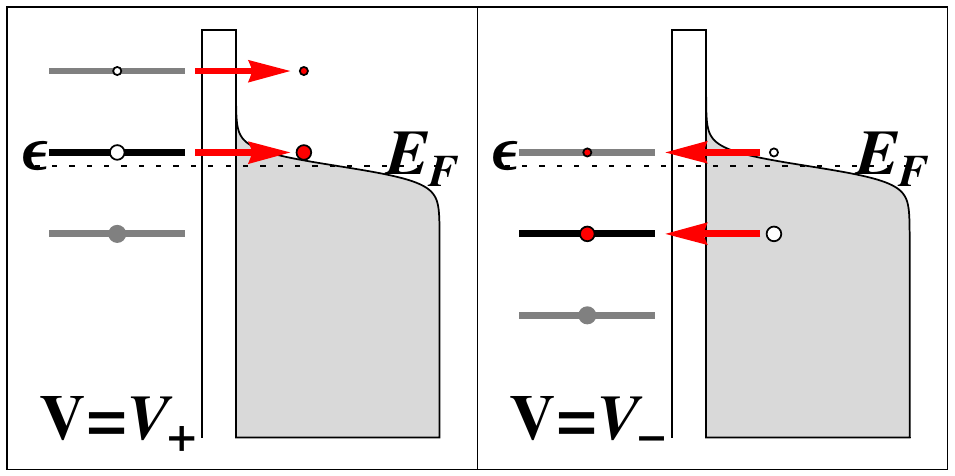}}
\includegraphics[width=\columnwidth]{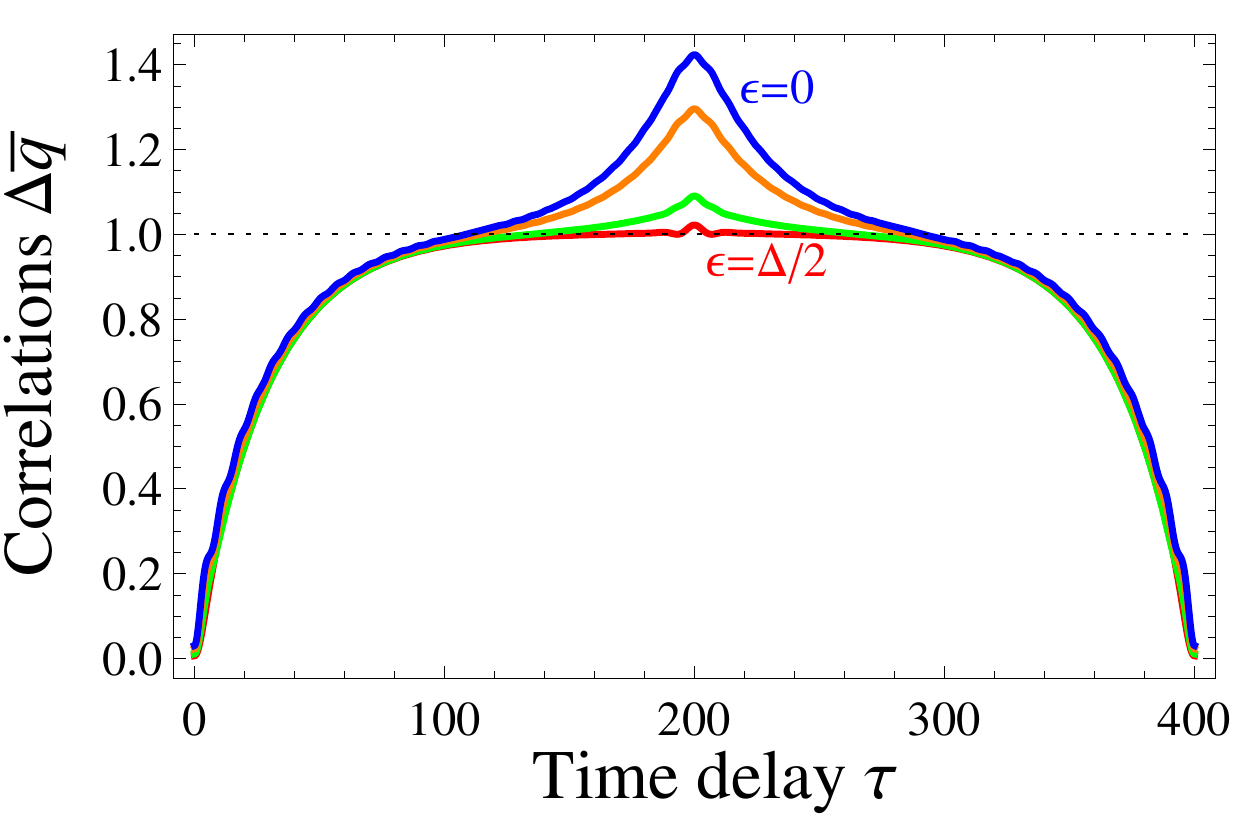}
\caption{\label{Fig:HOMElTr}
 Upper panel: Electron (left) and hole (right) emission process, for a square voltage drive of period $T_0=400$ (in units of $\hbar/\Delta$), for the two
positions of the dot levels (values $V_+$ and $V_-$ of the drive). The
position of the dot levels is parametrized by $\epsilon$ with respect to the Fermi energy $E_F$.
Bottom panel:  Theoretical prediction for the excess noise $\Delta \bar{q}$ as a function of the time delay $\tau$,
showing the electron-hole HOM peaks around $\tau=T_0/2=200$. The different curves are for $\epsilon = 0.5,0.4, 0.25$ and $0$ (in units of $\Delta$). $T_{el}=0.1 \Delta$ and the transparency $D=0.2$ in both panels.}
\end{figure}

To simulate the electron-hole HOM peak with the real electron emitters,
we have used the Floquet scattering matrix formalism. We have computed the correlations $\Delta \bar{q}$ when the two single electron sources
in the two input arms of the beam splitter are submitted to a square drive. As these sources periodically emit an electron and then
(after half a period) a hole, the correlations obtained for a time delay close to a half-period correspond to the correlations
between an electron and a hole. The results for $\Delta \bar{q}$ as a function of the time-delay $\tau$
 are shown on Fig.\ref{Fig:HOMElTr}, for a drive period of 400 (in units of $\hbar/\Delta$). As the correlations are proportional to the overlap
in energy of the electron and the hole wavefunctions (see Eq.(\ref{eq:HOM_eh})), in order to observe a peak
the electron emission and the hole emission need to happen
at energies not too far apart. This can be controlled by the dot level position of the single electron source
with respect to the Fermi energy: when a dot level is close to resonance with the Fermi energy ($\epsilon=0$ on Fig.\ref{Fig:HOMElTr}),
the energy overlap between the emitted electron and the emitted hole is important, and a large peak in the correlations $\Delta \bar{q}$
is observed. On the other hand, when the dot levels of the single electron sources are far from resonance ($\epsilon=0.5$ on
Fig.\ref{Fig:HOMElTr}), there is no overlap in energy between the emitted electron and the emitted hole, and
no peak is visible in the correlations, as observed on the experimental data of Fig. \ref{Fig:HOMDip} where electron/hole correlations are below experimental resolution. The temperature used in these simulations is $T_{el} = 0.1 \Delta$, which is similar to the experimental value.

\subsection{Tomography of a periodic electron source}

In the previous experiments, properties of the source can be inferred by measuring, through current correlations, the  resemblance between the state in input arm 1 and its counterpart in input arm 2. Indeed, HBT correlations yield information on the energy distribution of the source, by taking the Fermi sea as a reference, whereas HOM correlation demonstrate the indistinguishability of two quantum states generated by two independent sources. In fact, the complete coherence function in energy domain $\Delta \tilde{\mathcal{G}}^{(1,e)}(\epsilon,\epsilon')$ of a source of electrons and holes can be obtained in the HBT geometry by placing in input arm 2 different reference sources and measuring the corresponding current correlations. These spectroscopy \cite{Moskalets2011} and tomography processes \cite{Grenier2011NJP}, inspired by the optics equivalent \cite{Smithey1993,Bertet2002,Ourjoumtsev2006} could provide a direct image of electron wavepackets propagating in quantum Hall edge channels through the determination of the first order coherence in the $\epsilon$, $\epsilon'$ plane. For a periodic source, the definition of the first order coherence in the energy domain needs to be slightly modified. Indeed, $\Delta \mathcal{G}^{(1,e)}(t,t')$ has a T-periodicity in the time $\bar{t}=\frac{t+t'}{2}$, and no periodicity along $\tau=t-t'$. Using these two variables in time, the Fourier transform is defined in the following way:
\begin{eqnarray}
\mathcal{G}^{(1,e)}(t,t') = \sum_{n=-\infty}^{+ \infty} e^{-in \Omega \bar{t}} \int \frac{d\omega}{2 \pi} \tilde{\mathcal{G}}^{(1,e)}_n(\omega) e^{-i \omega \tau}
\end{eqnarray}
From the above definition, $\tilde{\mathcal{G}}^{(1,e)}_n(\omega)$ and $\tilde{\mathcal{G}}^{(1,e)}(\epsilon, \epsilon')$ are related through:
\begin{eqnarray}
\tilde{\mathcal{G}}^{(1,e)}(\epsilon, \epsilon') = \sum_{n=-\infty}^{+ \infty} \frac{\delta(\epsilon - \epsilon' - n\hbar \Omega)}{h} \tilde{\mathcal{G}}^{(1,e)}_n \Big(\frac{\epsilon+ \epsilon'}{2} \Big) \quad
\end{eqnarray}
Due to the periodicity in time, $\tilde{\mathcal{G}}^{(1,e)}(\epsilon, \epsilon')$ takes discrete values along the energy difference $\epsilon - \epsilon' = n \hbar \Omega$ while the sum of energies takes continuous values, $\frac{\epsilon + \epsilon'}{2} = \hbar \omega$. The population in energy domain thus corresponds to the $n=0$ component  of $\tilde{\mathcal{G}}^{(1,e)}_n(\omega)$ while the coherences correspond to $n \neq 0$.

\begin{figure}
  \includegraphics[width=\columnwidth]{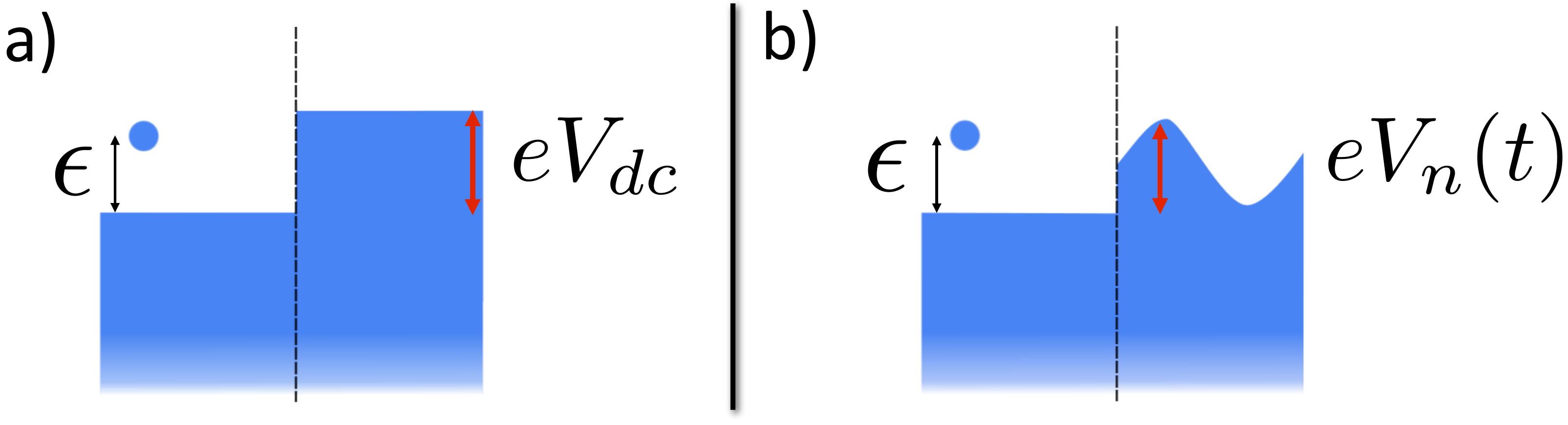}
  \caption{\label{Fig:SketchSpectroTomo}
   The spectroscopy and tomography of a periodic electron source can be achieved by modulating in a controlled way the two-particle interference, in the HBT geometry, between the source under study and reference sources. a) Sweeping the voltage $V_{dc}$ applied on the ohmic contact in input 2 enables to extract the diagonal part of the coherence function of the source in input 1, namely the energy distribution $\delta n_{e/h}$. b) A dynamical modulation of the partition noise by applying a voltage $V_n(t)=V_{dc}+V_{ac}\cos(n\Omega t+\phi)$ similarly gives access to the harmonics $\Delta \mathcal{G}_n, n\neq 0$ of the coherence function.}
\end{figure}

\begin{figure*}[t]
  \includegraphics[width=0.76\textwidth]{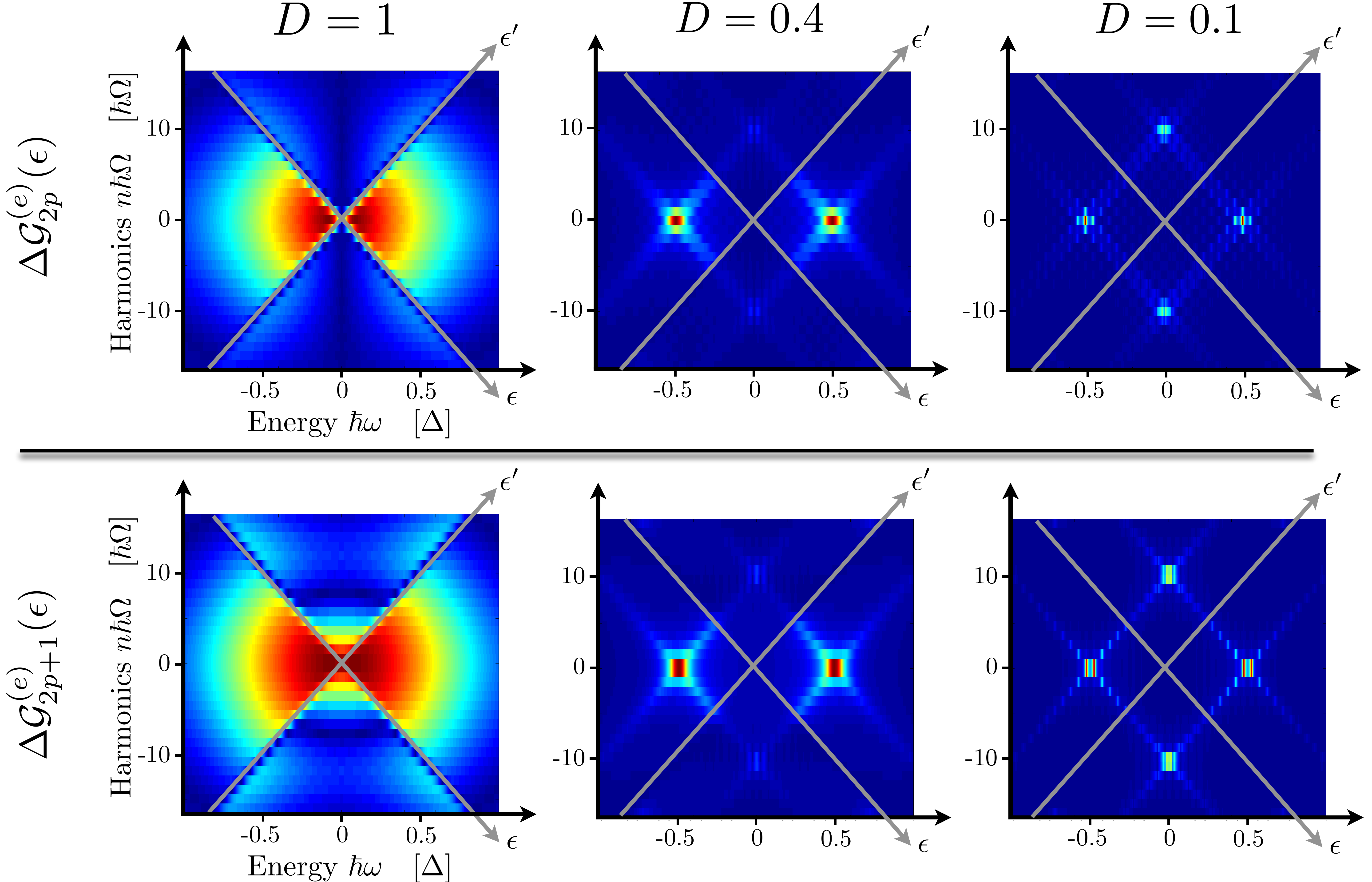}
  \caption{\label{Fig:TomoWavepackets} Examples of coherence functions in the complex plane. For transmissions $D=1$, $D=0.4$, $D=0.1$, odd and even harmonics of coherence functions $\Delta\mathcal{G}^{(e)}$ are plotted as a function of energy in a 2D plots. In contrast with the case $D=1$, excitations are energy resolved at rather high energies $\pm\Delta/2$ for $D=0.4$ and $D=0.1$. When emission probability drops (for $D=0.1$), emission of holes and electrons are correlated as the generation of an electron is subject to the generation of the preceding hole}
\end{figure*}

The source contribution of the coherence function $\Delta \tilde{\mathcal{G}}^{(1,e)}_n(\omega)$ can be fully reconstructed in the $n \hbar \Omega = \epsilon - \epsilon'$, $\hbar \omega = \frac{\epsilon + \epsilon'}{2}$ plane by applying as a reference state on input $2$ a voltage $V_n(t)=V_{dc}+V_{ac}\cos(n\Omega t+\phi)$ sum of a dc bias and an ac excitation at angular frequency $n\Omega$. The complete description of this tomography protocol lies beyond the scope of this article and can be found in Ref.\cite{Grenier2011NJP}. However an intuitive understanding can be drawn, that mainly relies on the two-particle interference between the electron source under study and the reference source. Let us first focus on the reconstruction of the $n=0$ component of the coherence function, associated with the energy distribution,  $\delta n_{e/h}$ that is on the spectroscopy of the electron source. A sketch supporting this discussion is presented Fig.\ref{Fig:SketchSpectroTomo} a). In the case $n=0$, only the dc part of the voltage applied on input $2$ is kept: $V_0(t)=V_{dc}$ that shifts the chemical potential of the connected edge by the value $-e V_{dc}$ .  As already mentioned, a two-particle interference can only occur between states of same energy. An electron at a well defined energy $\epsilon_0$ finds a symmetric partner in input 2 only if $\epsilon_0<-eV_{dc}$ (in the limit of vanishing temperature). Under this threshold, antibunching occurs with unit probability and partition noise is reduced to zero. Otherwise, for $\epsilon_0> -e V_{dc}$, the random partitioning takes place, regardless of the presence of the DC bias. Accordingly, by sweeping the bias $V_{dc}$, one can then reconstruct the probability of finding a particle at energy $\epsilon$, namely $\delta n_{e/h}(\epsilon) = \frac{h^2 f}{v} \Delta \tilde{\mathcal{G}}^{(1,e)}_n(\epsilon/\hbar)$ from the $V_{dc}$ dependence of the partition noise due to antibunching effects. Due to thermal smearing effects, the resolution of such a spectroscopy is in fact limited to $kT_{el}$ in the presence of a finite temperature $T_{el}$. In the same manner (Fig.\ref{Fig:SketchSpectroTomo} b)), dynamical modulations of the noise with a reference voltage $V_n(t)=V_{dc}+V_{ac}\cos(n\Omega t+\phi)$ enables to gain access to harmonics $\Delta \tilde{\mathcal{G}}^{(1,e)}_n(\omega)$ for $n\neq0$, that is the off diagonal elements $\epsilon -\epsilon' = n\hbar \Omega$ in the $\epsilon, \epsilon'$ plane.

Using once again the Floquet scattering formalism, simulations of the coherence function of a periodic source have been realized, in the case of the single electron/hole source: three cases with different sets of parameters illustrate the key features on Fig.\ref{Fig:TomoWavepackets}. For clarity, odd and even harmonics of $\Delta\mathcal{G}^{(e)}(\omega)$ are plotted on separate graphs as they have different parity with respect to $\omega$: $\Delta\mathcal{G}^{(e)}_{2p}$ is odd while $\Delta\mathcal{G}^{(e)}_{2p+1}$ is even. First, these graphs clearly highlight the four quadrants identified in Fig.\ref{fig:Quadrants}.
For a transmission $D=0.4$, the parameters are close to the optimal values : every charge is emitted during the dedicated emission cycle and the excitations are highly energy-resolved, around energies $\pm \Delta/2$. Only weak $e/h$ coherences are detected: the emission probability is very close to one, so that the emission of an electron is decorrelated from the emission of the previous hole as the emission probability is close to one.
Going towards higher transmission ($D=1$) yields excitations that lie mostly at low energy, and spread over a wide range of energies. Since the transmission is high, the two emission events of electron and holes are once again decorrelated. On the opposite, for lower transmissions ($D=0.1$), strong $e/h$ coherences appear as the emission probability is much smaller than 1. Production of holes and electrons are correlated as the emission of an electron is subject to the emission of the preceding hole, which does not take place in each cycle.

Note that, as suggested in refs \cite{Haack2011,Haack2013},  the coherence function of the source could also be measured in time domain,  $\Delta \mathcal{G}^{(1,e)}(t,t+\tau)$, measuring the current at time $t$ at the output of a Mach-Zehnder interferometer as a function of the difference $\tau$ in the propagation time between the two arms of the interferometer. This method implies a simpler measurement (average current instead of current fluctuations) but a more complicated sample. Also, decoherence effects during the propagation in the interferometer \cite{Sukhorukov2007, Levkivskyi2008,Levkivskyi2009,Chirolli2013} would have to be taken care of.

\section{Interactions in electron quantum optics}
\label{Interactions}

\subsection{Interaction mechanism in quantum Hall edge channels}
\label{Interactionmechanism}
In the previous sections of this manuscript, electron-electron interactions have been neglected, regarding the presentation of the general framework of electron quantum optics (section \ref{Formalism})  as well as in the discussion of the experimental results where the propagation along the channels was assumed to be
interaction free and dissipationless. Most results can indeed be first analyzed without taking into account the presence of interaction-induced decoherence of the mono-electronic excitations. However, due to their one-dimensional nature, quantum Hall edge channels are prone to emphasize interaction effects. In 1D systems, the motion of an electron interacting with its neighbours strongly affects the latter, so that the picture of quasi-free quasiparticles (Fermi liquid paradigm) holding for 2D and 3D systems is not adapted. It is replaced by the Luttinger liquid description, that relies on bosonic collective excitations \cite{Giamarchi2003}, called edge-magnetoplasmons in quantum Hall systems. Moreover, inter-channel Coulomb interactions then couple neighboring edge channels (at filling factor $\nu>1$), leading to the appearance of new collective propagation eigenmodes \cite{Levkivskyi2008}.

\begin{figure}[hhhh]
  \includegraphics[width=\columnwidth]{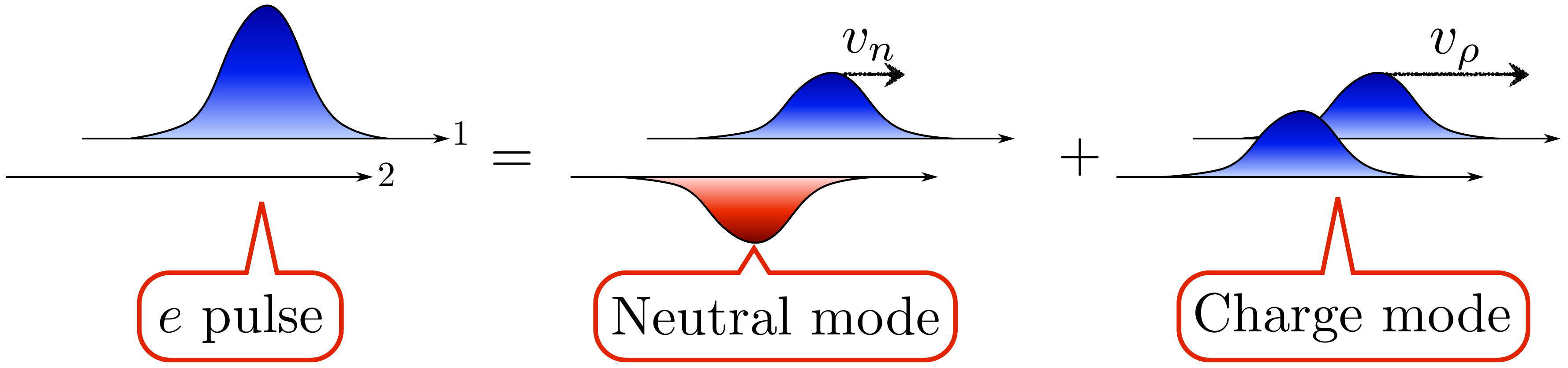}
  \caption{\label{Fig:ModeDecomposition}%\col
    In case of strong coupling between edge channels, a charge density wave in channel 1 is decomposed on two new propagation eigenmodes: a slow neutral mode of velocity $v_n$ with antisymmetric distribution of the charge, and a fast charge mode (velocity $v_\rho\gg v_n$) with a symmetric repartition of the charge.}
\end{figure}

The simplest model of two interacting co-propagating edge channels ($\nu=2$) illustrates the typical interaction mechanism. In the absence of both inter and intra channel interactions, currents propagate independently in each channel at the bare Fermi velocity $v$. The current $i_k(x, \omega)$ flowing in channel $k$ ($k=1, 2$) at position $x$ and angular frequency $\omega$ is simply related to the current at position $x=0$ by the phase $e^{i\frac{\omega x}{v}}$ acquired along the propagation: $i_k(x, \omega)=e^{i\frac{\omega x}{v}}i_k(0)$. If only intrachannel interactions are turned on, channels 1 and 2 are not coupled such that current propagation along each channel is still described by a phase with a velocity renormalized by interactions. However, when including interchannel interactions, outcoming currents $i_k(x, \omega)$ at position $x$ are related to incoming ones at position $x=0$ via a $2\times2$ scattering matrix $S_{emp}(\omega,x)$ \cite{Degiovanni2010}. Note that $S_{emp}$ describes the scattering of edge magnetoplasmons and not electrons, so that $S_{emp}$ acts on the current rather than on the fermion field operator $\hat{a}$ in usual Landauer-B\"{u}ttiker scattering formalism. The diagonalization of the scattering matrix $S_{emp}$ then gives access to the new propagation eigenmodes, that couple both channels. In particular, in the limit of strong interactions the two eigenmodes consist in a slow neutral dipolar mode for which the charge is anti-symmetrically distributed between both channels, and fast charge mode with symmetric charge distribution \cite{Levkivskyi2008}  as depicted in Fig.\ref{Fig:ModeDecomposition}. Due to Coulomb repulsion, the charge mode propagates much faster than the neutral one, $v_\rho\gg v_n$. The appearance of these eigenmodes bears strong similarities with the separation of the spin and charge degrees of freedom in non-chiral quantum wires \cite{Pham2000, Auslaender2005, Jompol2009}.

Various experiments have been carried out to investigate the coupling between edge channels and their effect on the relaxation and decoherence of electronic excitations. This coupling has been shown to be responsible for the loss of the visibility of the interference pattern in Mach-Zehnder interferometers at filling factor $\nu=2$  \cite{Neder2006, Roulleau2007, Roulleau2008}. In this case, the coupling of the external channel (which is the one probed in the interferometer) to the neighboring one leads to decoherence as information on the quantum state generated in the outer channel is capacitively transferred to the inner one acting as the environment. The influence of interchannel coupling on the energy relaxation of out of equilibrium excitations emitted in the outer edge channel has also been probed \cite{Altimiras2009, LeSueur2010} at filling factor $\nu=2$ using a quantum dot as an energy filter. These results have shown that coherence is lost and energy relaxes on a typical length of a few microns. Numerous theoretical works have successfully interpreted decoherence in interferometers \cite{Sukhorukov2007, Levkivskyi2008, Kovrizhin2010} and energy relaxation along propagation \cite{Lunde2010, Degiovanni2010, Kovrizhin2011, Levkivskyi2012} as stemming from interchannel Coulomb interactions. As a consequence, decoherence and relaxation can be controlled to some extent for example by the use of additional gates used to screen the interchannel interaction or by closing the internal edge channel which then acquires a gapped discrete spectrum such that interactions are fully frozen for energies below the gap. The latter technique has been shown to decrease both the energy relaxation \cite{Altimiras2010} and the coherence length \cite{Huynh, Huynh2012}.

Coupling between channels have also been investigated through high frequency current measurements that directly probe the propagation of edge magnetoplasmons in a quantum Hall circuit. Numerous experimental works have investigated the propagation of charge along quantum Hall edge channels, both in the time \cite{Ashoori1992,Zhitenev1993,Sukhodub2004} or in the frequency domain \cite{Gabellia,Talyanskii1992,Hashisaka2012}. However, in the $\nu=2$ case for example, to access all the terms of the $2\times2$ scattering matrix $S_{emp}(\omega,x)$ and reveal the nature of the eigenmodes, one needs to selectively address each edge channel individually. Using a mesoscopic capacitor to selectively inject an edge magnetosplasmon in the outer edge channel and a quantum point contact to analyze the scattering of the emp to the outer and inner edge channels after a controlled interaction length, the scattering parameters of $S_{emp}(\omega,x)$ and their frequency dependence could be investigated, thus revealing the nature of the neutral and charge eigenmodes by a direct measurement of the current at high frequency \cite{Bocquillon2013b}. Recently, interchannel interactions could also be characterized using partition noise measurements \cite{Inoue2013} to measure the excitations (electron/hole pairs) induced in the inner channel when electrons were injected selectively in the outer one.

The existence and nature of the interchannel coupling is thus now well established, however its influence on any arbitrary single electron state generated above the Fermi sea by single particle emitter is a challenging problem that still requires theoretical and experimental investigation. Some results can be obtained in the specific case of a single electron state emitted at a perfectly well defined energy $\epsilon_0$ above the Fermi sea \cite{Degiovanni2009}.

\subsection{Decoherence of an energy-resolved excitation}

\begin{figure}[h!]
\centering\includegraphics[width=\columnwidth]{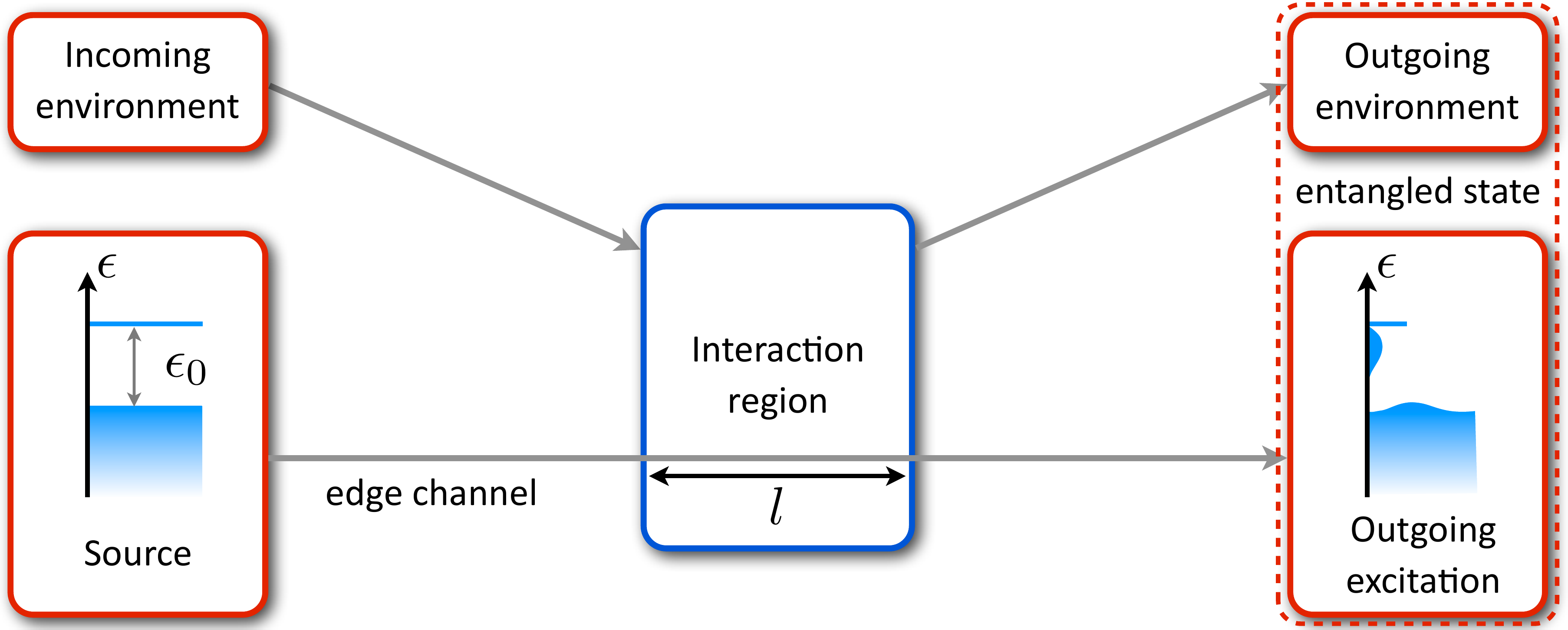}
\caption{Schematics of interactions with the environment. A single particle propagating on an edge channel enters the interaction region. After the emission of plasmonic waves in the environment, the edge channel is entangled with the environment at the output. }\label{fig5_5}
\end{figure}

  A single electronic excitation created on top of the Fermi sea enters at $x=0$ in a region where it interacts, via Coulomb interaction, with an environment along a propagation length $l$, see Fig. \ref{fig5_5}. The external environment, which can be any capacitively coupled conductor like an external gate or the adjacent edge channel in the $\nu=2$ case is labeled as conductor $2$ while the edge channel along which the excitation propagates will be labeled as conductor $1$. As discussed previously in the context of two coupled edge channels at $\nu=2$, the interaction between both conductors can be encoded in the scattering matrix $S_{emp}(\omega,l)$ which gives the scattering coefficients for charge density waves of angular frequency $\omega$ propagating in conductors $1$ and $2$ from the input $x=0$ to the output $x=l$ of the interaction region. During propagation in the interaction region, a single particle will emit plasmonic waves in the environment (in the following the input state of the environment will be considered to be at equilibrium at zero temperature). The environment and edge channel $1$ are then described by a complex many-body state where the edge channel and environment are entangled. Tracing out the environmental degrees of freedom at the output, the state of edge channel $1$ cannot be described as a pure state anymore and the off-diagonal terms of the first order coherence function can be drastically reduced.

\begin{figure}
  \includegraphics[width=\columnwidth]{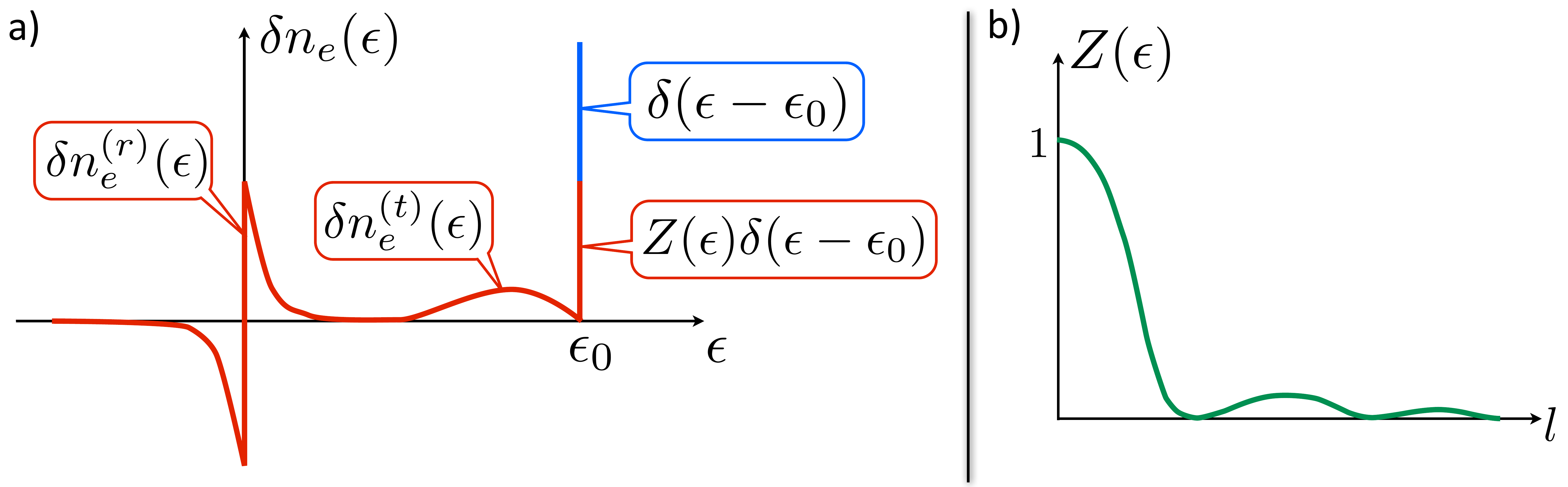}
  \caption{\label{Fig:RelaxMonochromatic}%\col
   a) Energy distribution before (blue curve) and after (red curve) interaction along the propagation length $l$. b) Typical dependence of the quasiparticle peak height on the interaction length $l$.}
\end{figure}

 The single electron coherence that describes the electronic state in edge channel 1 at the input of the interaction region ($x,y \geq 0$) is known, $\Delta \mathcal{G}^{(1,in)}(x,y) = \phi^{e}(x) \phi^{e,*}(y) \propto e^{i \frac{\epsilon_0 (x-y)}{\hbar v}}$ (we prefer here to use the $x, y$ notation than the $t, t'$ one to distinguish between the input and output of the interaction region).  In Fourier space, the energy distribution consists of a Dirac peak at energy $\epsilon_0$ above the Fermi sea, $\delta n_{e}(\epsilon) = \delta(\epsilon-\epsilon_0)$ (see blue curve on Fig.\ref{Fig:RelaxMonochromatic}.a)). Note that as a consequence of the specific choice of the input wavepacket (plane wave of well defined energy), the input state is stationary in time such that the coherence function in Fourier space is fully determined by the diagonal part $\delta n_{e}(\epsilon)$. At the output of the interaction region, one can guess the shape of the output energy distribution: the electron has lost some energy, as a consequence, the quasiparticle peak is reduced to the height $Z \leq 1$ (which eventually goes down to zero as the propagation length increases, see Fig.\ref{Fig:RelaxMonochromatic}.b)) and a relaxation tail $\delta n_{e}^{(t)}(\epsilon)$ appears below the quasiparticle peak. This energy can be transferred both to the environment but also to the Fermi sea through the creation of additional electron-hole pairs. This can be seen by the appearance of a non-equilibrium energy distribution $\delta n_{e}^{(r)}(\epsilon)$ at small energies above the Fermi sea. At high enough energy $\epsilon_0$ each of these two contributions can be identified and associated with a decoherence coefficient of the emitted wavepacket: $\phi^{e}(x) \phi^{e,*}(y) \rightarrow \phi^{e}(x) \phi^{e,*}(y) \; \mathcal{D}(x-y)$ with $ \mathcal{D}(x-y) =\mathcal{D}_{FS}(x-y) \times  \mathcal{D}_{env}(x-y) $ where $\mathcal{D}_{FS}$ stands for a Fermi sea induced decoherence and  $\mathcal{D}_{env}$ for the decoherence induced by the external environment. These two decoherence coefficients can be directly expressed as a function of the plasmon scattering matrix in the interaction region \cite{Degiovanni2009}:
\begin{eqnarray}
\mathcal{D}_{env}(x-y)=\exp{\int^\infty_0 \frac{d\omega}{\omega} |S_{21}(\omega)|^2 \big(e^{i\frac{\omega(x-y)}{v}}-1\big)} \quad \quad \quad \; \; \\
\mathcal{D}_{FS}(x-y)=\exp{\int^\infty_0 \frac{d\omega}{\omega}|1 - S_{11}(\omega)|^2 \big(e^{i\frac{\omega(x-y)}{v}}-1\big)}  \quad \quad \\
\mathcal{D}(x-y)=\exp{\int^\infty_0 \frac{d\omega}{\omega}2 \Re\big(1-S_{11}(\omega)\big)\big(e^{i\frac{\omega(x-y)}{v}}-1\big)} \quad \quad
\end{eqnarray}
In this regime, the Fermi sea appears as an extra dissipation channel which must be taken into
account into an effective environment. Note that here, this picture emerges in the high energy limit and is not valid when the extra-particle
relaxes down to the Fermi surface. In this latter case, separation of the extra particle and the additional electron-hole pairs created above the Fermi sea is not possible and the decoherence coefficient $\mathcal{D}(x-y)$ cannot be identified as easily.
This decoherence coefficient which suppresses the off diagonal coefficients of the first order coherence ($\mathcal{D}(x-y) \rightarrow 0 $ for $|x-y| \rightarrow \infty$) has important consequences on a Hong-Ou-Mandel experiment which is a sensitive probe of the off-diagonal components (coherences).
Let us assume for simplicity that the decoherence factor takes the simple form $\mathcal{D}(t,t') = e^{- \frac{|t-t'|}{\tau_c}}$ (the decoherence factor has been expressed in time instead of position using $x=-vt$).  In this case, Eq.(\ref{eq:HOM}) for the normalized output noise in the HOM experiment which was valid in the case of two pure states at the input of the splitter (absence of decoherence) needs by the following expression which takes into account decoherence:
\begin{eqnarray}
\Delta \overline q & =& 1 - \int dt dt' \phi^{e}_1(t) \phi^{e,*}_1(t') \phi^{e}_2(t') \phi^{e}_2(t) \mathcal{D}_1(t,t')\mathcal{D}_2(t',t) \nonumber \\
 \end{eqnarray}
 Taking $\phi^{e}_1(t) = \frac{\Theta(t)}{\sqrt{\tau_e}} e^{-\frac{t}{2 \tau_e}} e^{-i \epsilon_0 t/\hbar}$, $\phi^{e}_2(t) = \phi^{e}_1(t+\tau) $ where $\tau$ is the tuneable time delay between the emission of the two sources, and $\mathcal{D}_1(t,t') = \mathcal{D}_2(t,t')= e^{-\frac{|t-t'|}{\tau_c}}$, one obtains:
 \begin{eqnarray}
\Delta \overline q & =& 1 - \eta e^{-\frac{|\tau|}{\tau_e}} \\
\eta & =& \frac{1}{1+ 2 \tau_e/\tau_c}
 \end{eqnarray}

This model of decoherence predicts a reduction of the HOM dip at $\tau=0$ while keeping the shape of an exponential decay when varying the delay $\tau$. This model predicts that a wavepacket with a small temporal extension and in particular much smaller than the coherence time $\tau_c$ is not affected by decoherence, $\eta \approx 1$. On the contrary, a wavepacket with a large temporal extension ($\tau_e \gg \tau_c$) is drastically affected by decoherence and the HOM dip vanishes, $\eta \approx \frac{\tau_c}{2 \tau_e}$. In this limit, the electron cannot be described by a coherent wavepacket with a well defined phase relationship between its various temporal component but rather by a classical probability distribution of different emission times of typical extension given by $\tau_e$. In this case, the width $\tau_e$ plays the role of a random delay between the two sources which explains the reduction of the HOM dip. In our experiment, we measure $\eta \approx 0.5$ for $\tau_e \approx 50$ ps which is consistent with $\tau_c \approx 100$ ps.

\subsection{Interactions in the Hong-Ou-Mandel setup}
We now provide a quantitative description of the effects of Coulomb interactions in the Hong-Ou-Mandel setup in the case of interchannel coupling at filling factor $2$. We consider the case of short range interchannel interactions and strong coupling such that the eigenmodes are the symmetric fast charge mode (with velocity $v_{\rho}$) and the slow antisymmetric neutral mode (with velocity $v_n \ll v_{\rho}$) as described in Sec.~\ref{Interactionmechanism}. Finite temperature of the leads can also be included.

The single electron source is modeled through the injection of single wave-packets at a given distance $l$ from the QPC (chosen symmetrically for the two incoming arms: $x=\pm l$). As previously discussed, the wavepackets are defined as exponentials in real-space, $\phi_2 (x) = \frac{1}{\sqrt{v \tau_e}} e^{-i \epsilon_0 x/(\hbar v)} e^{-x/(2 v \tau_e)} \theta(x)$, and for the sake of simplicity, we focus on the interference between identical wave-packets, $\phi_1 (x)=\phi_2 (-x)$.

The normalized HOM correlation then reads \cite{Wahl2013}:
	\begin{align}
		\Delta \bar{q} (\tau) = 1 - \frac{ \mathrm{Re} \left[ q_{HOM} \right]}{\mathrm{Re} \left[ q_{HBT} \right]} \label{DeltaqCPT}
	\end{align}
where
\begin{align}
 q_{HOM} &=  \int d x_1 d y_1 \int d x_2 d y_2 \phi_1 (x_1) \phi_1^* (y_1)  g(0,x_1-y_1)  \nonumber\\
		      & \times \phi_2 (x_2) \phi_2^* (y_2)  g (0,y_2-x_2)   \int d t  d t' \mathrm{Re} \left[ g(t' - t,0)^2 \right]  \nonumber\\
      & \times  \left[ 1 - \frac{h (t;x_2,y_2)}{h (t';x_2,y_2)} \right] \times \left[ 1 - \frac{h (t'+\tau;-x_1,-y_1)}{h (t+\tau;-x_1,-y_1)}  \right] \\
 q_{HBT} &= \int d x_1 d y_1 \int d x_2 d y_2 \phi_1 (x_1) \phi_1^* (y_1)  g(0,x_1-y_1)  \nonumber\\
		      & \times \phi_2 (x_2) \phi_2^* (y_2)  g (0,y_2-x_2)   \int d t  d t' \mathrm{Re} \left[ g(t' - t,0)^2 \right]  \nonumber\\
      & \times  \left[ 2 - \frac{h (t;x_2,y_2)}{h (t';x_2,y_2)} - \frac{h (t';-x_1,-y_1)}{h (t;-x_1,-y_1)} \right]
\end{align}
and the auxiliary functions introduced are given by
\begin{align*}
g(t,x)&= \left[ \frac{\sinh \left( i \frac{\pi a}{\beta v_\rho} \right)}{\sinh \left(  \frac{ia + v_\rho t - x}{\beta v_\rho /\pi} \right)}  \frac{\sinh \left( i \frac{\pi a}{\beta v_n} \right)}{\sinh \left( \frac{ia + v_n t - x}{\beta v_n /\pi} \right)} \right]^{1/2} , \\
h (t;x,y) &= \left[ \frac{\sinh \left( \frac{ia - v_\rho t + x + l}{\beta v_\rho / \pi} \right)}{\sinh \left(\frac{ia + v_\rho t - y - l}{\beta v_\rho/\pi} \right)} \right]^{\frac{1}{2}} \left[ \frac{\sinh \left( \frac{ia - v_n t + x + l}{\beta v_n / \pi} \right)}{\sinh \left( \frac{ia + v_n t - y - l}{\beta v_n /\pi} \right)} \right]^{\frac{1}{2}} .
\end{align*}
The variable $a$ is a spatial cutoff, which ultimately needs to be sent to 0, and $\beta = 1/(k_B T_{el})$.

\begin{figure}
\includegraphics[scale=0.43]{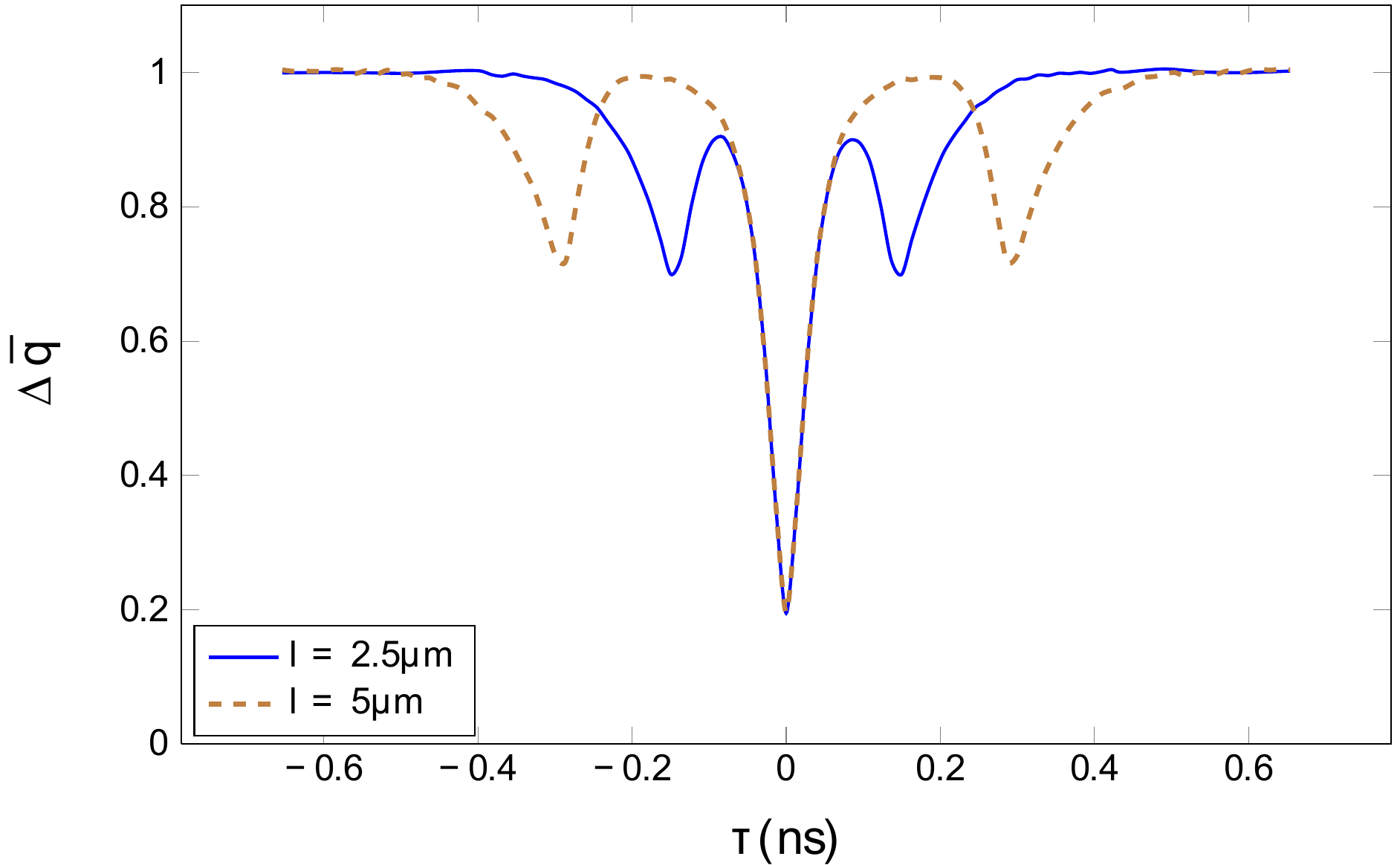}
\includegraphics[scale=0.43]{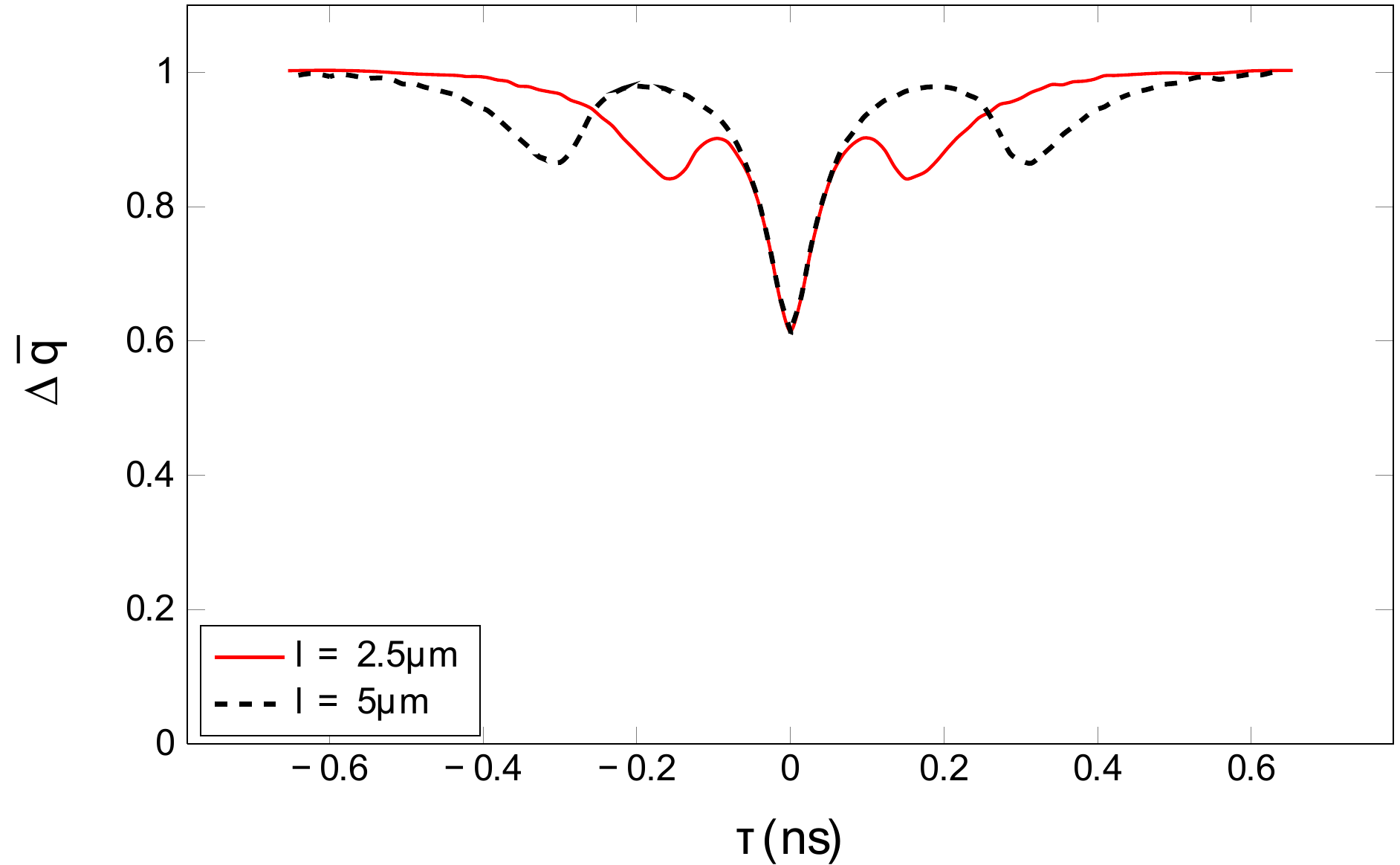}
\caption{Normalized HOM correlations as a function of the time delay $\tau$, for two different type of wave-packets: (upper) one with an escape time $\tau_e= 22$ ps and emitted energy $\epsilon_0=0.175$ K
and (lower) one with an escape time $\tau_e= 44$ ps and emitted energy $\epsilon_0=0.7$ K. In both cases, $T_{el} = 0.1$ K.
	}
\label{Fig:analyticHOM}
\end{figure}

Numerical evaluation of Eq. \ref{DeltaqCPT} can be performed thanks to a quasi Monte Carlo algorithm using importance sampling~\cite{Hahn2005}, results are presented on Fig.\ref{Fig:analyticHOM}.
As we vary the time delay $\tau$ of the right-moving electron over the left-moving one,
our computations reveal the presence of three characteristic signatures in the noise (see Fig.~\ref{Fig:analyticHOM}) : a central dip at $\tau = 0$, and two side structures which emerge symmetrically with respect to the central dip at $\tau = \pm l (v_\rho-v_n) /v_\rho v_n$. The depth and shape of these three dips are conditioned by the energy resolution of the incoming wave-packets. Away from these three features, the normalized correlations saturate at a constant value, representing the Hanbury-Brown and Twiss contribution. This corresponds to the situation where the electrons injected on the two incoming arms scatter independently at the QPC.

This interference pattern can be interpreted in terms of the different excitations propagating along the partitioned edge channel.  After injection, the electron fractionalizes into two modes: a slow neutral mode with anti-symmetric distribution of the charge between the injection and the co-propagating channels and a fast charge mode with a symmetric repartition of the charge among the two channels.
The central dip, which corresponds to the symmetric situation of synchronized injections, thus probes the interference of excitations with the same velocity and charge. These identical excitations interfere destructively, leading to a reduction of the noise (in absolute value), thus producing a dip in the normalized HOM correlations.

A striking difference with the non-interacting case is that the central dip never reaches down to $0$ as observed experimentally (see Sec.~\ref{HBThom}). The depth of this dip is actually a probing tool of the degree of indistinguishability between the colliding excitations~\cite{Feve2008}. Our present work suggests that because of the strong inter-channel coupling, some coherence is lost in the other channels, and the Coulomb-induced decoherence leads to this characteristic loss of contrast for the HOM dip.  This effect gets more pronounced for further energy-resolved packets. As depicted in Fig.~\ref{Fig:analyticHOM}, while for ``wide'' packets in energy ($\gamma=\frac{2\epsilon_0 \tau_e}{\hbar} \approx 1$) the contrast (defined as $\eta=1-\Delta \bar{q} (0)$) is still pretty good, $\eta \sim 0.8$, the loss of contrast can be dramatic for energy-resolved packets, with $\eta \sim 0.4$ for $\gamma=8$.

Adjusting the time delay appropriately, one can also probe interferences between excitations that have different velocities. This effect is responsible for the side structures appearing in the noise: at $\tau = l (v_\rho - v_n)/(v_\rho v_n)$, the fast right-moving excitation and the slow left-moving one reach the QPC at the same time while the dip at $\tau =-l (v_\rho - v_n)/(v_\rho v_n)$ corresponds to the collision between a slow right-moving excitation and a fast left-moving one. Like the central dip, these lateral structures correspond to the collision of two excitations of the same charge, which interfere destructively. Their depth is however less than half the one of the central dip. This can be attributed to the velocity mismatch between interfering excitations, as it indicates that they are more distinguishable. This difference of velocity of the two colliding objects is also responsible for the asymmetry of the lateral dips. Typically, the slope is steeper for smaller $| \tau |$.
This asymmetry is very similar to the one encountered in the non-interacting case for interfering packets with different shapes, where a broad right-moving packet in space collides onto a thin left-moving one~\cite{Jonckheere2012}.

\section{Conclusion}

As detailed in this manuscript, optical tools and concepts can be used in a very efficient way to understand and characterize electronic propagation in a quantum conductor. Within this framework, electronic transport is analyzed through a simple single particle description which captures most of the features of electron propagation but is only correct in the non-interacting photon-like case. In the presence of Coulomb interactions the correct description relies on the resolution of a complex many-body problem.

The production and manipulation of single-particle states provide a direct test bench for single-particle physics. Using controlled emitters with tuneable parameters, a wide range of single particle wavefunctions can be engineered both in time or energy \cite{Bocquillon2012, Mirovsky2013} space. Coulomb interaction during propagation with the surrounding electrons of the Fermi sea and nearby conductors will strongly affect the state of a single excitation. Consequently, even the propagation of a single electron tends to a complex many body problem : as the electronic wavepacket propagates, it relaxes and decoheres, and additional electron-hole excitations are generated. This mechanism sets the limits of electron quantum optics : during propagation, a single-particle excitation is diluted in collective excitations, so that the possibility of manipulating a pure single-particle quantum state is lost.
To get a complete understanding of the effects of Coulomb interactions, it is necessary to picture fully the electronic wavefunction in energy or time domains.  The tomography protocol suggested in \cite{Grenier2011} provides a complete imaging of the first order coherence in energy domain from noise measurements in the Hanbury-Brown and Twiss geometry. In particular the energy distribution of mono-electronic excitations could be extracted from the variation of the output noise when shifting the chemical potential of a Fermi sea used as reference state in one input. The measured energy distribution after a tuneable propagation length could be compared with the non interacting theory in analogy with the spectroscopy of a non-equilibrium stationary electron beam performed in ref.  \cite{Altimiras2009, Battista2012} using a quantum dot as an energy filter. The energy distribution is also directly related to heat transfers and heat fluctuations generated by single particle emitters  \cite{Moskalets2009, Battista2013} and could thus be inferred from nano-caloritronic measurements.
In the time domain, the first order coherence could be measured using a single electron emitter at the input of a Mach-Zehnder interferometer \cite{Haack2011}.

Beyond the study of the propagation of a single excitation, proposals have been made to manipulate coherently single to few electronic excitations, connecting the physics of quantum conductors to quantum information processing. For example, the Mach-Zehnder geometry, together with two single electron emitters placed at the input, could be used to postselect entangled electron pairs \cite{Splettstoesser2009,Sherkunov2012,Vyshnevyy2012} or to generate GHZ states \cite{Vyshnevyy2013}. However, such coherent manipulations would require to reduce and circumvent the effect of Coulomb interaction in quantum Hall edge channels for example by closing the internal edge channel \cite{Altimiras2010,Huynh2012}. Energy exchanges between neighboring edge channels are then frozen for energies below the excitation gap of the internal edge.  As pioneered in \cite{Karmakar2011}, coherent manipulations could also be performed on the spin degree of freedom. By transferring charge in a controlled manner between the two co-propagating edge channels of opposite spins at filling factor $\nu=2$, any coherent superpositions of spins could be achieved.

Finally, another extremely interesting route would be to extend these concepts to other ballistic electronic systems. Of particular interest would be the study of triggered charge emission along the edge channels of the fractional quantum Hall regime \cite{Jonckheere2005}. The question is whether one can emit and manipulate a single quasiparticle of fractional charge in the same fashion as single electronic excitations for integer values of the filling factor. In particular the study of two-particle interference would be of particular interest as they are sensitive to the phase associated with the exchange of two particles and could thus provide a way to measure the statistics of fractional excitations.  Another possible implementation would be the recently discovered helical edge states of quantum spin Hall effect \cite{Roth2009,Brune2012} : an equivalent of the mesoscopic capacitor in such a system has already been proposed \cite{Hofer2013,Inhofer2013}, enabling the generation of time-bin entangled pairs of electrons.

\noindent \textbf{Acknowledgement.} This work was supported by the ANR grant '1shot', ANR-2010-BLANC-0412. We warmly thank M. Albert, C. Flindt, G. Haack and M. Moskalets for fruitful discussions and Markus B\"{u}ttiker for his strong support and inspiration to this work.

\bibliographystyle{andp2012}
\bibliography{BiblioAdPFeve}

\end{document}